\documentclass[12pt,preprint]{aastex}

\begin{document}

\def\wisk#1{\ifmmode{#1}\else{$#1$}\fi}

\def\lt     {\wisk{<}}
\def\gt     {\wisk{>}}
\def\le     {\wisk{_<\atop^=}}
\def\ge     {\wisk{_>\atop^=}}
\def\lsim   {\wisk{_<\atop^{\sim}}}
\def\gsim   {\wisk{_>\atop^{\sim}}}
\def\kms    {\wisk{{\rm ~km~s^{-1}}}}
\def\Lsun   {\wisk{{\rm L_\odot}}}
\def\Zsun   {\wisk{{\rm Z_\odot}}}
\def\Msun   {\wisk{{\rm M_\odot}}}
\def\um     {$\mu$m}
\def\mic     {\mu{\rm m}}
\def\sig    {\wisk{\sigma}}
\def\etal   {{\sl et~al.\ }}
\def\eg     {{\it e.g.\ }}
 \def\ie     {{\it i.e.\ }}
\def\bsl    {\wisk{\backslash}}
\def\by     {\wisk{\times}}
\def\half {\wisk{\frac{1}{2}}}
\def\third {\wisk{\frac{1}{3}}}
\def\nwm2sr {\wisk{\rm nW/m^2/sr\ }}
\def\nw2m4sr {\wisk{\rm nW^2/m^4/sr\ }}

\title{A measurement of large-scale peculiar velocities of clusters of galaxies: technical details}

\author{A. Kashlinsky\altaffilmark{1,5}, F.  Atrio-Barandela\altaffilmark{2},
  D. Kocevski\altaffilmark{3}, H.  Ebeling\altaffilmark{4}}
\altaffiltext{1}{SSAI and Observational Cosmology Laboratory, Code
665, Goddard
  Space Flight Center, Greenbelt MD 20771} \altaffiltext{2}{Fisica Teorica,
  University of Salamanca, 37008 Salamanca, Spain} \altaffiltext{3}{Department
  of Physics, University of California at Davis, 1 Shields Avenue, Davis, CA
  95616} \altaffiltext{4}{Institute for Astronomy, University of Hawaii, 2680
  Woodlawn Drive, Honolulu, HI 96822} \altaffiltext{5}{e--mail:
  alexander.kashlinsky@nasa.gov}
\begin{abstract}
This paper presents detailed analysis of large-scale peculiar
motions derived from a sample of $\sim 700$ X-ray clusters and
cosmic microwave background (CMB) data obtained with WMAP. We use
the kinematic Sunyaev-Zeldovich (KSZ) effect combining it into a
cumulative statistic which preserves the bulk motion component
with the noise integrated down. Such statistic is the dipole of
CMB temperature fluctuations evaluated over the pixels of the
cluster catalog (Kashlinsky \& Atrio-Barandela 2000). To remove
the cosmological CMB fluctuations the maps are filtered with a
Wiener-type filter in each of the eight WMAP channels (Q, V, W)
which have negligible foreground component. Our findings are as
follows: The thermal SZ (TSZ) component of the clusters is
described well by the Navarro-Frenk-White profile expected if the
hot gas traces the dark matter in the cluster potential wells.
Such gas has X-ray temperature decreasing rapidly towards the
cluster outskirts, which we demonstrate results in the decrease of
the TSZ component as the aperture is increased to encompass the
cluster outskirts. We then detect a statistically significant
dipole in the CMB pixels at cluster positions. Arising exclusively
at the cluster pixels this dipole cannot originate from the
foreground or instrument noise emissions and must be produced by
the CMB photons which interacted with the hot intracluster gas via
the SZ effect. The dipole remains as the monopole component, due
to the TSZ effect, vanishes within the small statistical noise out
to the maximal aperture where we still detect the TSZ component.
We demonstrate with simulations that the mask and cross-talk
effects are small for our catalog and contribute negligibly to the
measurements. The measured dipole thus arises from the KSZ effect
produced by the coherent large scale bulk flow motion. The
cosmological implications of the measurements are discussed by us
in Kashlinsky et al (2008).
\end{abstract}

\section{Introduction}

In the popular gravitational instability picture for growth of the
large scale structure in the Universe, peculiar velocities on
large cosmological scales probe directly the peculiar
gravitational potential and provide important information on the
underlying mass distribution in the Universe [e.g. see review by
Kashlinsky \& Jones 1991]. Previous attempts to measure the
peculiar flows in the local Universe mostly used empirically
established (but not well understood theoretically) galaxy
distance indicators. While very important, such methods are
subject to many systematic uncertainties [e.g. see reviews by
\cite{strauss-willick,willick}] and lead to widely different
results.

Early measurements by \cite{rubin-ford} indicated large peculiar
flows of $\sim$700 km/sec. A major advance was made using the
``Fundamental Plane" (FP) relation for elliptical galaxies
\cite{7s-di,djorgovski} with the implication that elliptical
galaxies within $\sim 60h^{-1}$Mpc were streaming at $\sim 600$
km/sec with respect to the rest frame defined by the cosmic
microwave background (CMB) \cite{7s-motion}. Mathewson et al
(1992) used the Tully-Fisher (TF) relation for a large sample of
spiral galaxies suggesting that the flow of amplitude 600 km/sec
does not converge until scales much larger than $\sim 60 h^{-1}$
Mpc. This finding was in agreement with a later analysis by
\cite{willick99}. Employing brightest cluster galaxies as distance
indicators \cite{lauer-postman} measured a bulk flow of $\sim$700
km/sec for a sample 119 rich clusters of galaxies on scale of
$\sim$150$h^{-1}$Mpc suggesting significantly larger amount of
power than expected in the concordance $\Lambda$CDM model.
However, a re-analysis of these data \cite{hudson-ebeling} taking
into account the correlation between the luminosities of
brightest-cluster galaxies and that of their host cluster found a
bulk flow in a greatly different direction and at a smaller
amplitude. Using the FP relation for early type galaxies in 56
clusters \cite{hudson} find a bulk flow of a similarly large
amplitude of $\sim 630$ km/sec to \cite{lauer-postman} on a
comparable scale, but in a different direction. On the other hand,
a sample of 24 SNIa shows no evidence of significant bulk flows
out to $\sim 100 h^{-1}$ Mpc \cite{riess} and similar conclusion
is reached with the TF based survey of spiral galaxies by
\cite{courteau}. The directions associated with each bulk-flow
measurement are equally discrepant.

The current situation with measurements based on the various
distance indicators is confusing and it is important to find
alternative ways to measure the large scale peculiar flows. One
way to achieve this is via the kinematic component of the Sunyaev
Zeldovich (SZ) effect produced on the CMB photons from the hot
X-ray emitting gas in clusters of galaxies ([see review by
\cite{birkinshaw}]. The kinematic SZ (KSZ) effect is independent
of redshift and measures the line-of-sight peculiar velocity of a
cluster in its own frame of reference. For each individual cluster
the KSZ temperature distortion will be small and difficult to
measure. Attempts at measuring the peculiar velocities of
individual clusters from the KSZ effect using the current
generation of instruments lead to uncertainties of $\gsim 1000$
km/sec per cluster [see review by \cite{carlstrom}]. On the other
hand, as proposed by \cite{kab} (hereafter KA-B) for many clusters
moving at a coherent bulk flow one can construct a measurable
quantity using data on CMB temperature anisotropies which will be
dominated by the bulk flow KSZ component, whereas the various
other contributions will integrate down. This quantity, {\it the
dipole of the cumulative CMB temperature field evaluated at
cluster positions}, is used in this investigation on the 3-year
WMAP data in conjunction with a large sample of X-ray clusters of
galaxies to set the strongest to-date limits on bulk flows out to
scales $\sim 300 h^{-1}$Mpc.

In the accompanying Letter (Kashlinsky et al 2008) we summed the
results and their cosmological implications. These are obtained
using the KA-B method applied to 3-year WMAP CMB data and the
largest all-sky X-ray cluster catalog to date. This paper provides
the details relevant for the measurement and is structured as
follows: Sec \ref{steps} summarizes the KA-B method and the steps
leading to the measurement. Sec. \ref{catalog} describes the
cluster X-ray catalog used in this study and Sec. \ref{cmb}
outlines the CMB data processing. Sec. \ref{errors} discusses the
methods to estimate the errors followed by Sec \ref{results} with
the results on the dipole measurement. Sec. \ref{tsz} shows why
the measured dipole arises from the KSZ component due to the
cluster motion and Sec. \ref{calibration} dicusses the translation
of the measured dipole in $\mu$K into velocity in km/sec and its
uncertainty. Future prospects foreseeable at this time to improve
this measurement are discussed in Sec. \ref{future}. We summarize
our results in Sec. \ref{summary}.

\section{KA-B method and steps to the measurement}
\label{steps}

If a cluster at angular position $\vec{y}$ has the line-of-sight
velocity $v$ with respect to the CMB, the SZ CMB fluctuation at
frequency $\nu$ at this position will be $\delta_\nu(\vec
y)=\delta_{\rm TSZ}(\vec y)G(\nu)+ \delta_{\rm KSZ}(\vec
y)H(\nu)$, with $ \delta_{\rm TSZ}$=$\tau T_{\rm X}/T_{\rm e,ann}$
and $\delta_{\rm KSZ}$=$\tau v/c$. Here $G(\nu)\simeq-1.85$ to
$-1.35$ and $H(\nu)\simeq 1$ over the range of frequencies probed
by the WMAP data, $\tau$ is the projected optical depth due to
Compton scattering, $T_{\rm X}$ is the cluster electron
temperature and $k_{\rm B}T_{\rm e,ann}$=511 KeV. If averaged over
many isotropically distributed clusters moving at a significant
bulk flow with respect to the CMB, the kinematic term may dominate
enabling a measurement of $V_{\rm bulk}$. Thus KA-B suggested
measuring the dipole component of $\delta_\nu(\vec y)$. Below we
use the notation for $C_{1,{\rm kin}}$ normalized so that a
coherent motion at velocity $V_{\rm bulk}$ would lead to
$C_{1,{\rm kin}}= T_{\rm CMB}^2 \langle \tau \rangle^2 V_{\rm
bulk}^2/c^2$, where $T_{\rm CMB} =2.725$K is the present-day CMB
temperature. For reference, $\sqrt{C_{1,{\rm kin}}}\simeq 1
(\langle \tau \rangle/10^{-3}) (V_{\rm bulk}/100{\rm km/sec}) \;
\mu$K. When computed from the total of $N_{\rm cl}$ positions the
dipole also will have positive contributions from 1) the
instrument noise, 2) the thermal SZ (TSZ) component, 3) the
cosmological CMB fluctuation component arising from the
last-scattering surface, and 4) the various foreground components
at the WMAP frequency range. The latter contribution can be
significant at the two lowest frequency WMAP channels (K \& Ka)
and, hence, we restrict this analysis to the WMAP Channels Q, V \&
W which have negligible foreground contributions.

For $N_{\rm cl}\gg1$ the dipole of the observed $\delta_\nu$
becomes:
\begin{equation}
a_{1m} \simeq a_{1m}^{\rm KSZ} +a_{1m}^{\rm TSZ} + a_{1m}^{\rm
CMB} + \frac{\sigma_{\rm noise}}{\sqrt{N_{\rm cl}}}
\label{eq:dipole}
\end{equation}
Here $a_{1m}^{\rm CMB}$ is the residual dipole produced at the
cluster pixels by the primordial CMB anisotropies. The amplitude
of the dipole power is $C_1= \sum_{m=-1}^{m=1} |a_{1m}|^2$.

Additional contributions to eq. \ref{eq:dipole} come from
non-linear evolution/collapse of clusters (Rees \& Sciama 1968),
gravitational lensing by clusters (Kashlinsky 1988), unresolved
strong radio sources (present, for instance, in WMAP 5 year data,
Nolta et al 2008) and the Integrated Sachs-Wolfe effect from the
cluster pixels. All these effects have a dipole signal only when
clusters are inhomogenously distributed on the sky and is in turn
bounded from above by the amplitude of the monopole. The magnitude
of these contributions is at most $\sim 10\mu$K$^2$ in power (see
Aghanim, Majumdar \& Silk 2008 for a review on secondary
anisotropies) a factor of 10 smaller than the Thermal
Sunyaev-Zeldovich monopole amplitude. Furthermore, as we discuss
below, we find a dipole signal when the monopole vanishes, so our
measurements can not be significantly affected by all these
effects.

In the following sections we detail out the process that enabled
us to isolate the KSZ term in eq. \ref{eq:dipole}. The steps
leading to this measurement were:

$\bullet$ An all-sky catalog of X-ray selected galaxy clusters was
constructed using available X-ray data extending to $z\simeq 0.3$.

$\bullet$ The cosmological CMB component was removed from the WMAP
data using the Wiener-type filter with the best-fit cosmological
model.

$\bullet$ The filter is constructed (and is different) for each DA
channel because the beam and the noise levels are different. This
then prevents inconsistencies and systematic errors that could
have been generated if a common filter was applied to the eight
channels of different noise and resolution.

$\bullet$ The filtered CMB maps were used to compute the dipole
component at the cluster positions simultaneously as the TSZ
monopole vanishes because of the X-ray temperature decrease with
radius (Atrio-Barandela et al 2008 and below).

$\bullet$ Simulations showed that the measured dipole arises from
the cluster pixels at a high confidence level. Since the TSZ
component from the clusters vanishes, only a contribution from the
KSZ component, due to large-scale bulk motion of the cluster
sample, remains.

The following sections present the technical details related to
this analysis.

\section{X-ray data and catalogue}
\label{catalog}

The creation of the all-sky cluster catalogue used here from three
independent X-ray selected cluster samples is described in detail
by Kocevski \& Ebeling (2006); for clarity we briefly reiterate
the procedure in the following.

The REFLEX catalog consists of 447 clusters with X-ray fluxes
greater than $3\times 10^{-12}$ erg cm$^{-2}$ s$^{-1}$ in the
[0.1--2.4] KeV band.  The survey is limited to declinations of
$\delta < 2.5^{\circ}$, redshifts of $z\leq 0.3$ and Galactic
latitudes away from the Galactic plane ($|b|>20^{\circ}$). The
eBCS catalog comprises 290 clusters in the Northern hemisphere
with X-ray fluxes greater than $3\times 10^{-12}$ erg cm$^{-2}$
s$^{-1}$ [0.1--2.4] KeV at Galactic latitude $|b| > 20^\circ$. The
sample is limited to declinations of $\delta > 0^{\circ}$ and
redshifts of $z\leq 0.3$ and, like REFLEX, the survey avoids the
Galactic plane ($|b|>20^{\circ}$). The CIZA sample is the product
of the first systematic search for X-ray luminous clusters behind
the plane of the Galaxy.  The sample contains 165 clusters with
X-ray fluxes greater than $3\times 10^{-12}$ erg cm$^{-2}$
s$^{-1}$ [0.1--2.4] KeV and redshifts of $z\leq 0.3$.

To obtain a single homogeneous sample the physical properties of
all clusters were recalculated in a consistent manner using
publicly available RASS data. Cluster positions were redetermined
from the centroid of each system's X-ray emission and point
sources within the detection aperture are removed. Total X-ray
count rates within an aperture of 1.5 $h_{50}^{-1}$Mpc radius were
calculated taking into account the local RASS exposure time and
background, and converted into unabsorbed X-ray fluxes in the
ROSAT broad band [0.1--2.4] KeV. Total rest-frame luminosities
were determined from the fluxes using the cosmological luminosity
distance and a temperature-dependent \emph{K}-correction.  Finally
clusters whose X-ray emission appeared to be dominated by a point
source were removed and a flux cut was applied at $3\times
10^{-12}$ erg cm$^{-2}$ s$^{-1}$, leaving 349 REFLEX, 268 eBCS,
and 165 CIZA clusters at $z\leq 0.3$.  The resulting sample is the
largest homogeneous, all-sky, X-ray selected cluster catalog
compiled to date, containing 782 clusters over the entire sky.  Of
these, 468 fall within $z\le0.1$. Further details concerning the
statistical properties of the catalog, including its completeness,
can be found in Kocevski \& Ebeling (2006). Figure 1 shows the sky
distribution of the clusters used in this analysis.

Our analysis requires knowledge of several parameters describing
the properties of the intra-cluster gas. We determine the X-ray
extent of each cluster directly from the RASS imaging data using a
growth-curve analysis. The cumulative profile of the net count
rate is constructed for each system by measuring the counts in
successively larger circular apertures centered on the X-ray
emission and subtracting an appropriately scaled X-ray background.
The latter is determined in an annulus from 2 to 3 $h_{\rm
50}^{-1}$ Mpc around the cluster centroid.  The extent of each
system is then defined as the radius at which the increase in the
source signal is less than the $1\sigma$ Poissonian noise of the
net count rate.  This is essentially the distance from the cluster
center at which the X-ray emission is no longer detectable with
any statistically significance.

\clearpage
 \begin{figure}[h]
\plotone{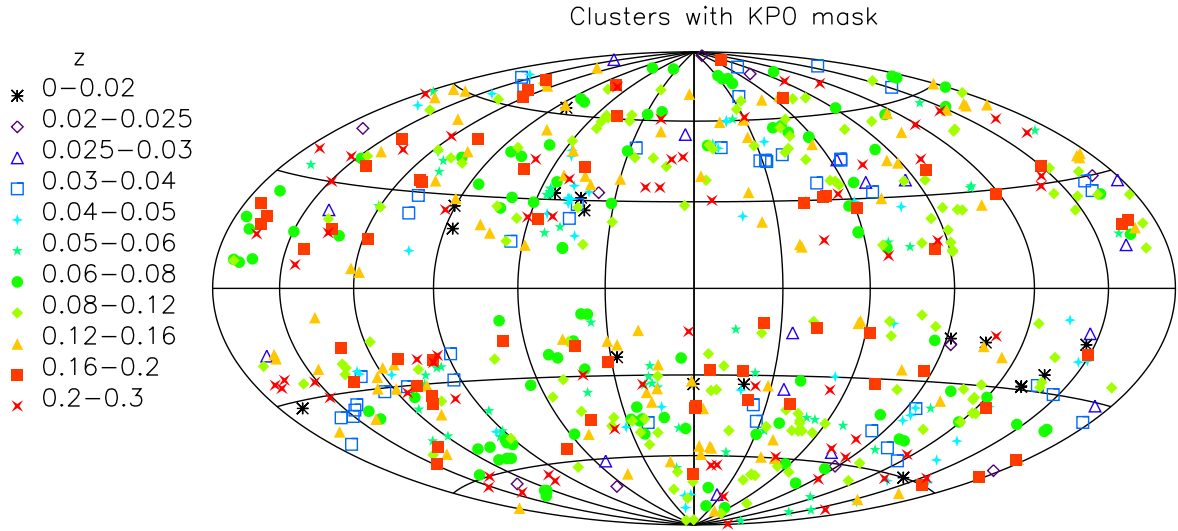}
  \caption{X-ray catalogue used in the paper with the KP0 mask applied.
  Note that at the lowest $z$ clusters have significant N:S asymmetry (for $z\leq 0.02,
0.025,0.03, 0.04$ there are 11:6, 16:11, 24:19, 44:42 N:S
clusters), which goes away at $z\gsim 0.03$.\label{fig:xcat}}
\end{figure}
\clearpage

Unabsorbed cluster fluxes were determined from our recalculated
count rates by folding the ROSAT instrument response against the
predicted X-ray emission from a Raymond-Smith (Raymond \& Smith
1977) thermal plasma spectrum with 0.3 solar metallicity and by
taking into account Galactic absorption in the direction of the
source. The temperature used in the spectral model is determined
iteratively using the cluster redshift, a first-order
approximation on the cluster luminosity using $k_{\rm B} T_{\rm
X}=4$ KeV and the $L_{\rm X}-T_{\rm X}$ relation of White et al
(1997). Total rest-frame $[0.1-2.4]$ KeV band cluster luminosities
were subsequently determined from our recalculated fluxes using
the standard conversion with the cosmological luminosity distance
and a temperature dependent $K$-correction.

To obtain an analytic parametrization of the spatial profile of
the X-ray emitting gas and, ultimately, the central electron
density we fit a $\beta$ model (Cavaliere \& Fusco-Femiano 1976)
convolved with the RASS point-spread function to the RASS data for
each cluster in our sample: $S(r) = S_0
\left[1+(r/r_c)^2\right]^{-3\beta+1/2}$ where $S(r)$ is the
projected surface-brightness distribution and $S_0$, $r_c$, and
$\beta$ are the central surface brightness, the core radius, and
the $\beta$ parameter characterizing the profile at large radii.
Using the results from this model fit to determine the gas-density
profile assumes the gas to be isothermal and spherically
symmetric.  In practice, additional uncertainties are introduced
by the correlation between $r_c$ and $\beta$ which makes the
results for both parameters sensitive to the choice of radius over
which the model is fit, and the fact that for all but the most
nearby clusters the angular resolution of the RASS allows only a
very poor sampling of the surface-brightness profile (at $z> 0.2$
the X-ray signal from a typical cluster is only detected in
perhaps a dozen RASS image pixels). In recognition of these
limitations, we hold $\beta$ fixed at the canonical value of $2/3$
and only allow $r_c$ to vary (Jones \& Forman 1984).  As a
consistency check, we also calculate $r_c$ values from each
cluster's X-ray luminosity using the $r_{c} \propto L_{\rm
X}^{1/3.6}$ empirical relationship determined by Reiprich \&
B\"{o}hringer (1999).  Our best-fit values for $r_c$ are
reassuringly robust in the sense that we find broad agreement with
the empirically derived values.

Our best-fit parameters, the cluster luminosity and electron
temperature, are used to determine central electron densities for
each cluster using Equation 6 of Henry \& Henriksen (1986) with
the temperature of the ICM being estimated from the $L_{\rm X} -
T_{\rm X}$ relationship of White, Jones, \& Forman (1997).  The
electron densities are in turn used to translate the CMB dipole in
$\mu$K into an amplitude in km/sec as described below.  We also
calculated electron densities using our empirically derived
cluster parameters and find good agreement between the resulting
dipole amplitude and the amplitude obtained using our best-fit
values.

 \clearpage
\begin{figure}[h]
 \plotone{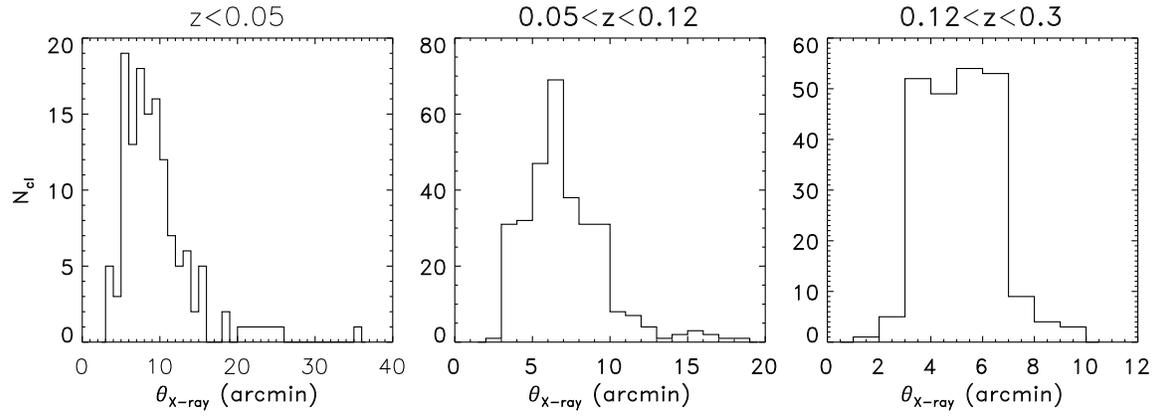}
 \caption{Distribution of cluster
X-ray extent in various $z$-bins using the KP0 maps. Coma is the
only cluster with X-ray radial extent larger than 0.5 deg.\label{fig:theta_x}}
\end{figure}
\clearpage

The distribution of the cluster radial extents determined by the
X-ray emission, $\theta_{\rm X-ray}$, for our catalog is shown in
Fig. \ref{fig:theta_x}.  Coma at $z\simeq 0.02$ has the largest
extent $\theta_{\rm X-ray} \simeq 35^\prime$. In order to avoid
the few large clusters, such as Coma, bias the determination of
the dipole, we introduce a cutoff of $30^\prime$ in the net extent
when increasing the size to account for the extent of the
SZ-producing gas. The final analysis was made increasing cluster
X-ray extent to $6\theta_{\rm X-ray}$ and then cutting them at
30$^\prime$ to ensure robust dipole computation. In the process
the variations in the cluster size across the sky become greatly
reduced: e.g. for the entire sample of 674 clusters which survive
the KP0 CMB mask, the final mean radial extent of the clusters is
$28.4^\prime$, standard deviation is $3.2^\prime$ and only 16
clusters have radii below $20^\prime$. Thus in our final
measurements all clusters are effectively 30 arcmin in radius
independently of the cluster position.

Conversions between angular extents and the physical dimensions of
clusters are made using the concordance cosmology
($\Omega_\Lambda=0.7, \Omega_{\rm total}=1, h=0.7$).

\section{CMB data processing and filtering}
\label{cmb}

\begin{figure}[h]
\plotone{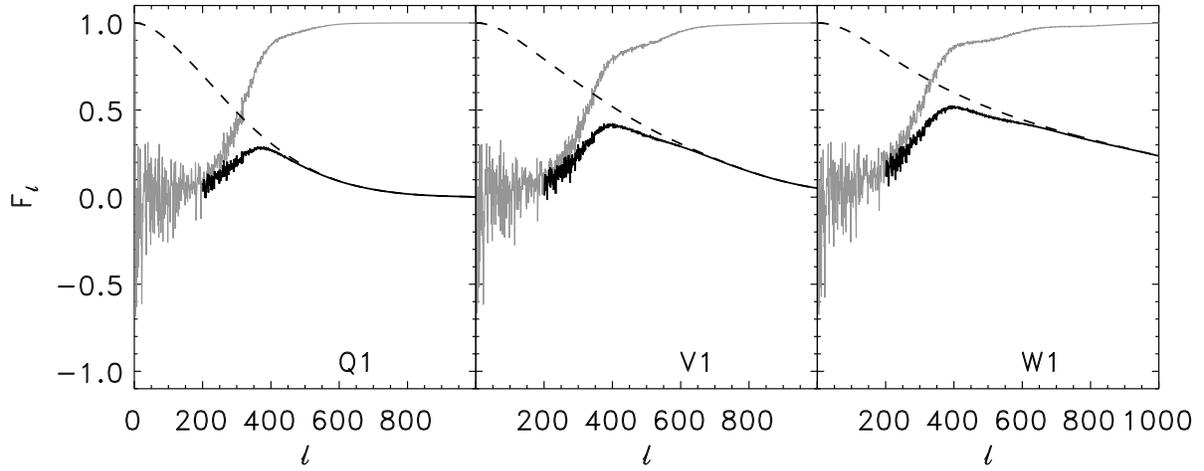}
 \caption{Filters used in removing the cosmological CMB fluctuations
 are shown with light-shaded lines. Dashed lines show the beam profiles for the marked WMAP channels.
 Solid lines show the product of the two: $B_\ell F_\ell$.\label{fig:filters}}
\end{figure}

Our starting point are the 3-year WMAP ``foreground-cleaned" maps
available from
http://lambda.gsfc.nasa.gov/product/map/current/m\_products.cfm in
two Q channels (Q1, Q2), two V channels (V1 and V2), and four W
channels (W1 through W4).  Channels K and Ka contain fairly
significant foreground emission and are not considered in this
study. Each channel has its own noise of variance $\sigma_n^2$
with the Q channels having the lowest noise and the W channels the
highest. The beam transfer functions for each channel, $B_\ell$,
were obtained from the same URL. The beam is also different in
each channel with Q1 having the poorest resolution and W4 the
highest. Examples of the beam profile are shown in Fig.
\ref{fig:filters}. The maps were masked of foreground emitters
using the KP2 and KP0 masks.

The resolution of the input maps is set by choosing $N_{\rm
side}=512$ in HEALpix (Gorski et al 2005). This corresponds to
pixels of $4\times 10^{-6}$ sr (47.2 arcmin$^2$) in area or
$\theta_p \simeq 6.87^\prime$ on the side. This resolution is much
coarser than that of the X-ray maps used for constructing our
cluster catalog.

\clearpage
\clearpage

Because cosmological CMB fluctuations are correlated, they could
leave a significant variance in the noise component of our
measurement (eq. \ref{eq:dipole}) over the relatively few pixels
occupied by the clusters. Of course, this noise component will be
the same, within its standard deviation, for any other pixels in
the maps, rather than being peculiar to the cluster pixels.
Because the power spectrum of this component, $C_\ell^{\rm \Lambda
CDM}$, is accurately known from WMAP studies (Hinshaw et al 2007),
it can be effectively filtered out of the CMB maps, substantially
reducing its contribution to the noise budget in eq.
\ref{eq:dipole}. This can be achieved with the Wiener filter,
which minimizes the mean square deviation from the noise $\langle
(\delta T - \delta_{\rm noise})^2\rangle$ (e.g. Press et al 1986).
The Fourier transform of this filter is:
\begin{equation}
F_\ell = \frac{C_\ell({\rm sky}) - C_\ell^{\rm \Lambda CDM}
B_\ell^2}{C_\ell({\rm sky})}
 \label{eq:filter}
\end{equation}
where $C_\ell({\rm sky})$ is the Fourier transform of the sky
which contains both the $\Lambda$CDM component and the instrument
noise.

\clearpage
\begin{figure}[h]
 \begin{tabular}{cc}
 \includegraphics[width=1.625in,angle=90]{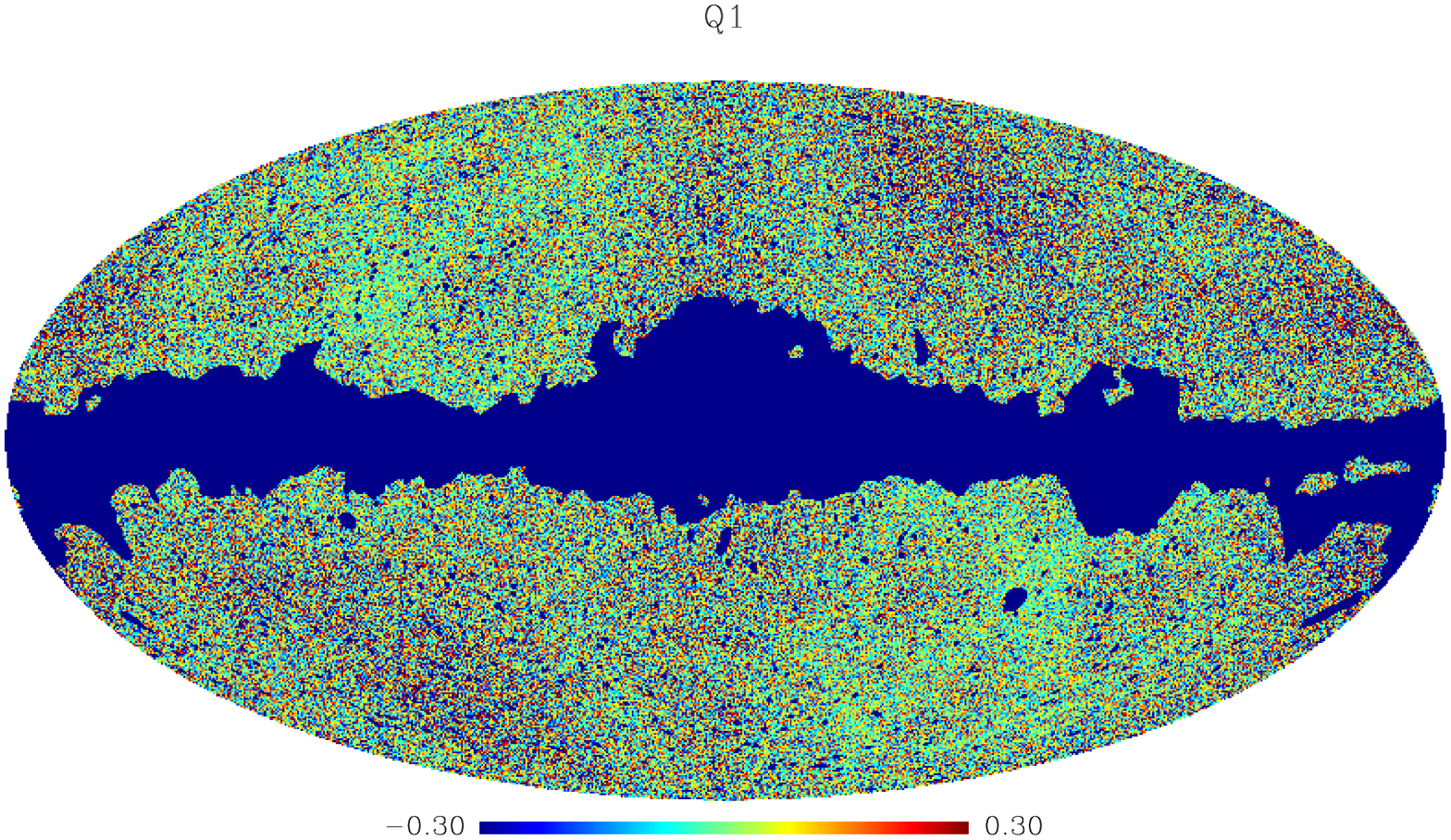} & \includegraphics[width=1.625in,angle=90]{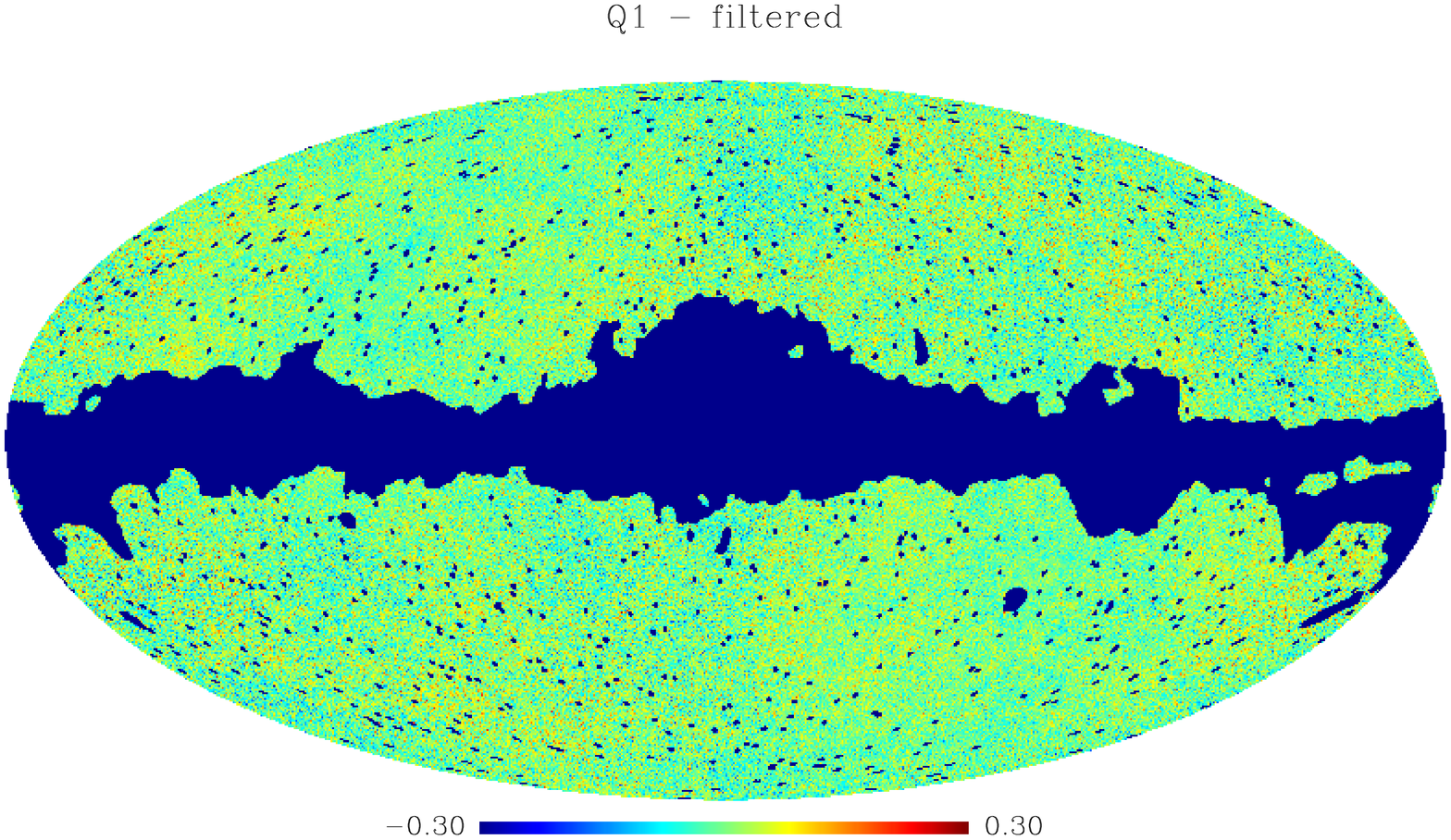} \\
 \includegraphics[width=1.625in,angle=90]{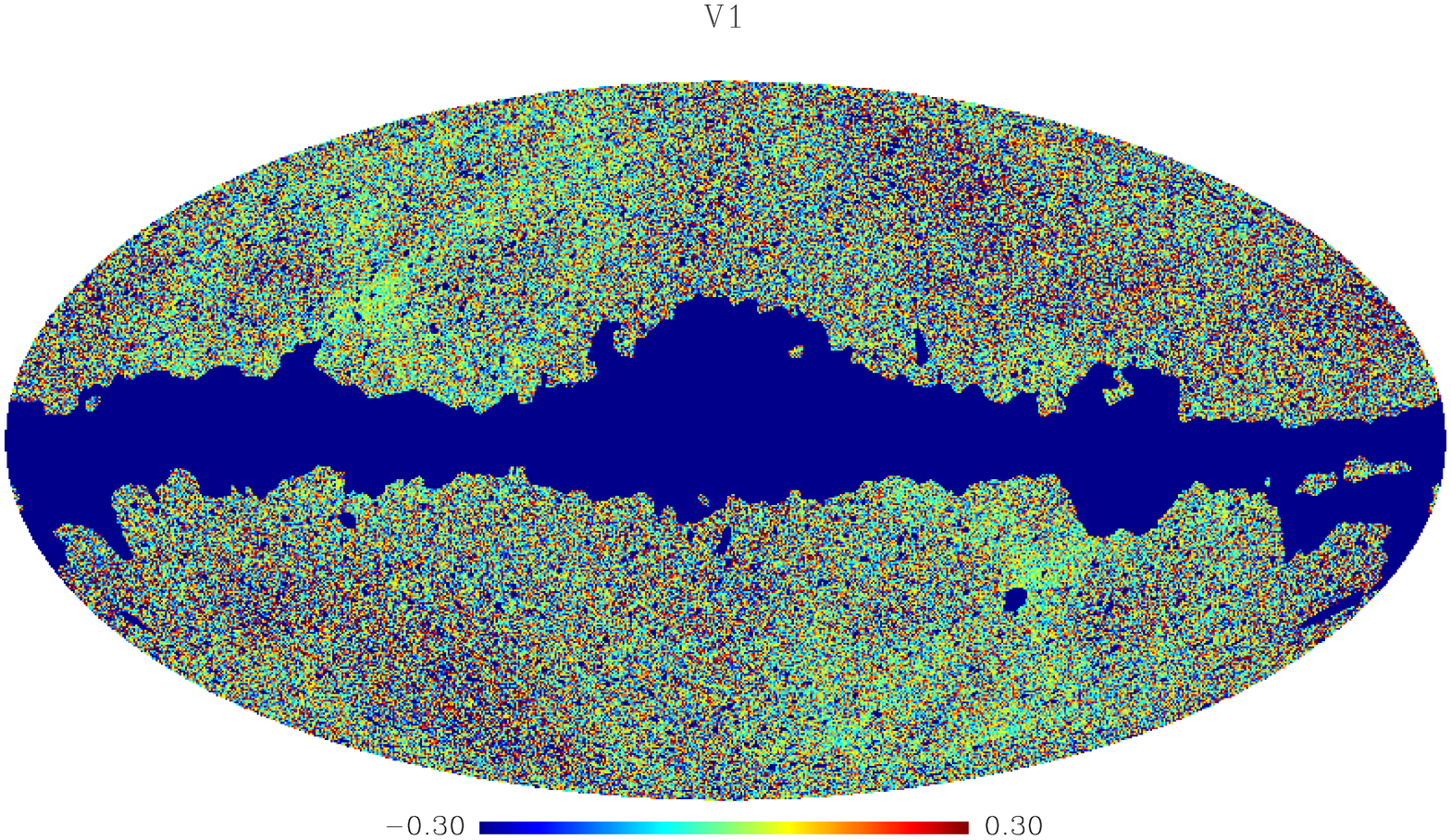} & \includegraphics[width=1.625in,angle=90]{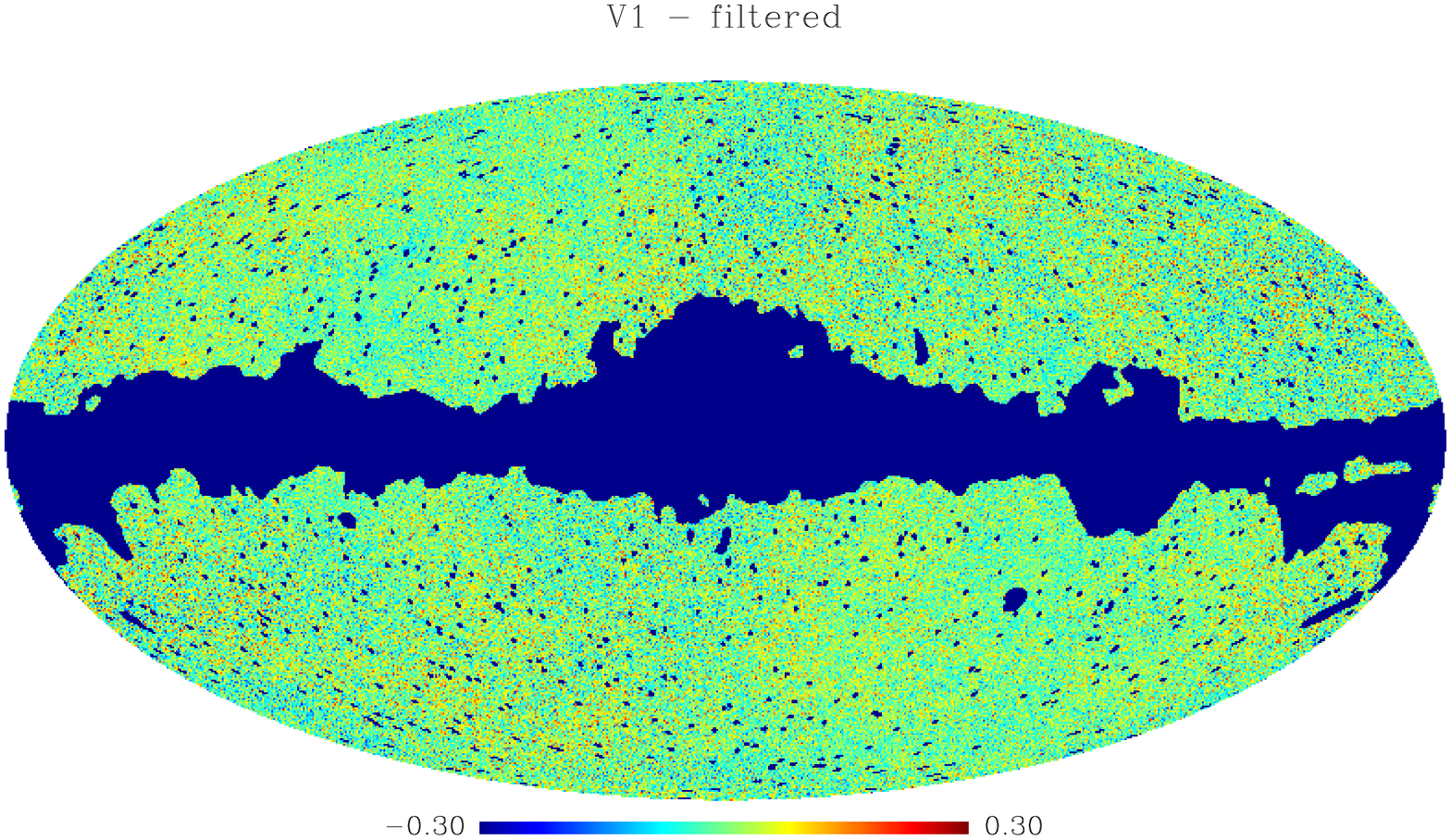} \\
 \includegraphics[width=1.625in,angle=90]{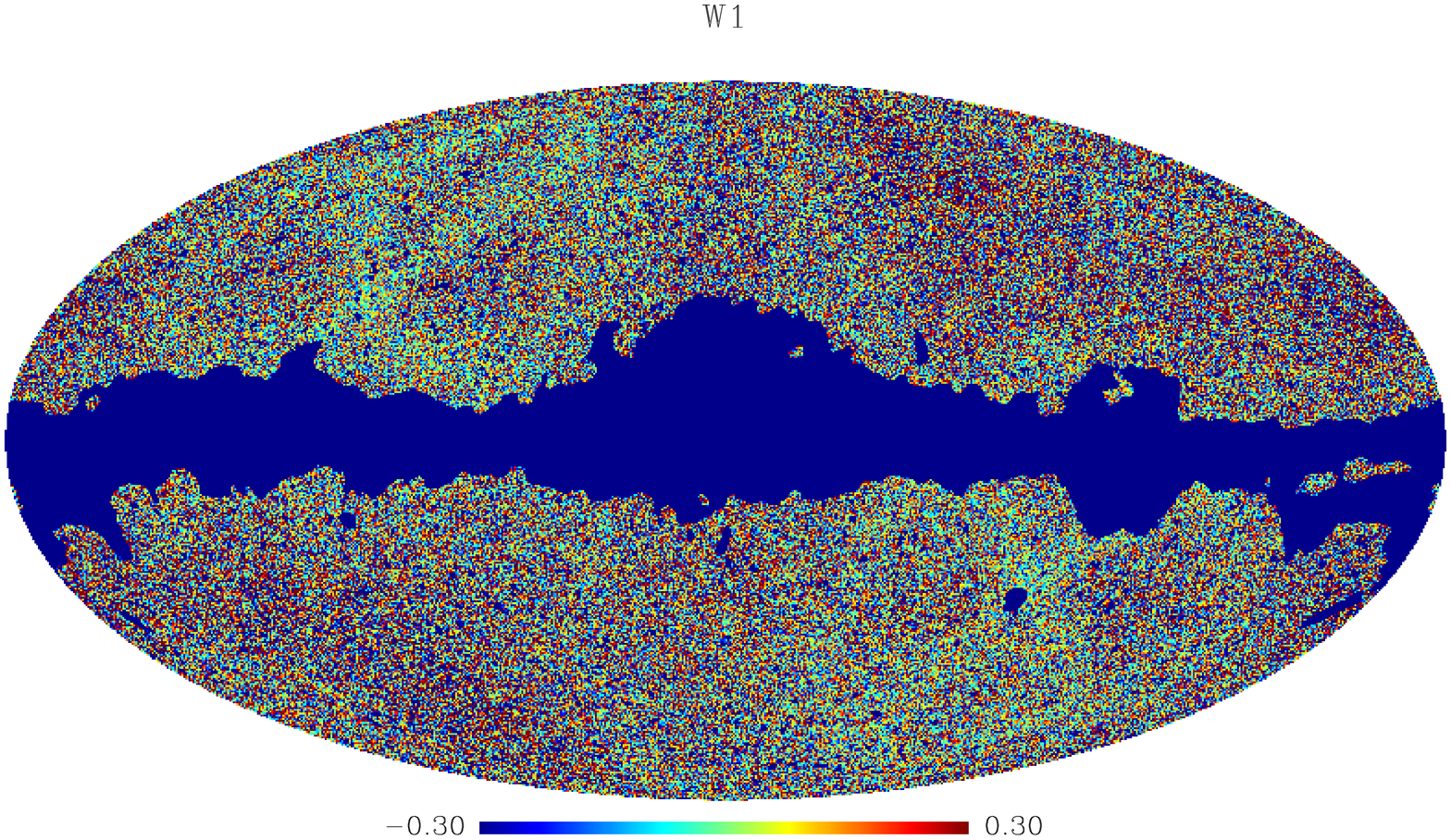} & \includegraphics[width=1.625in,angle=90]{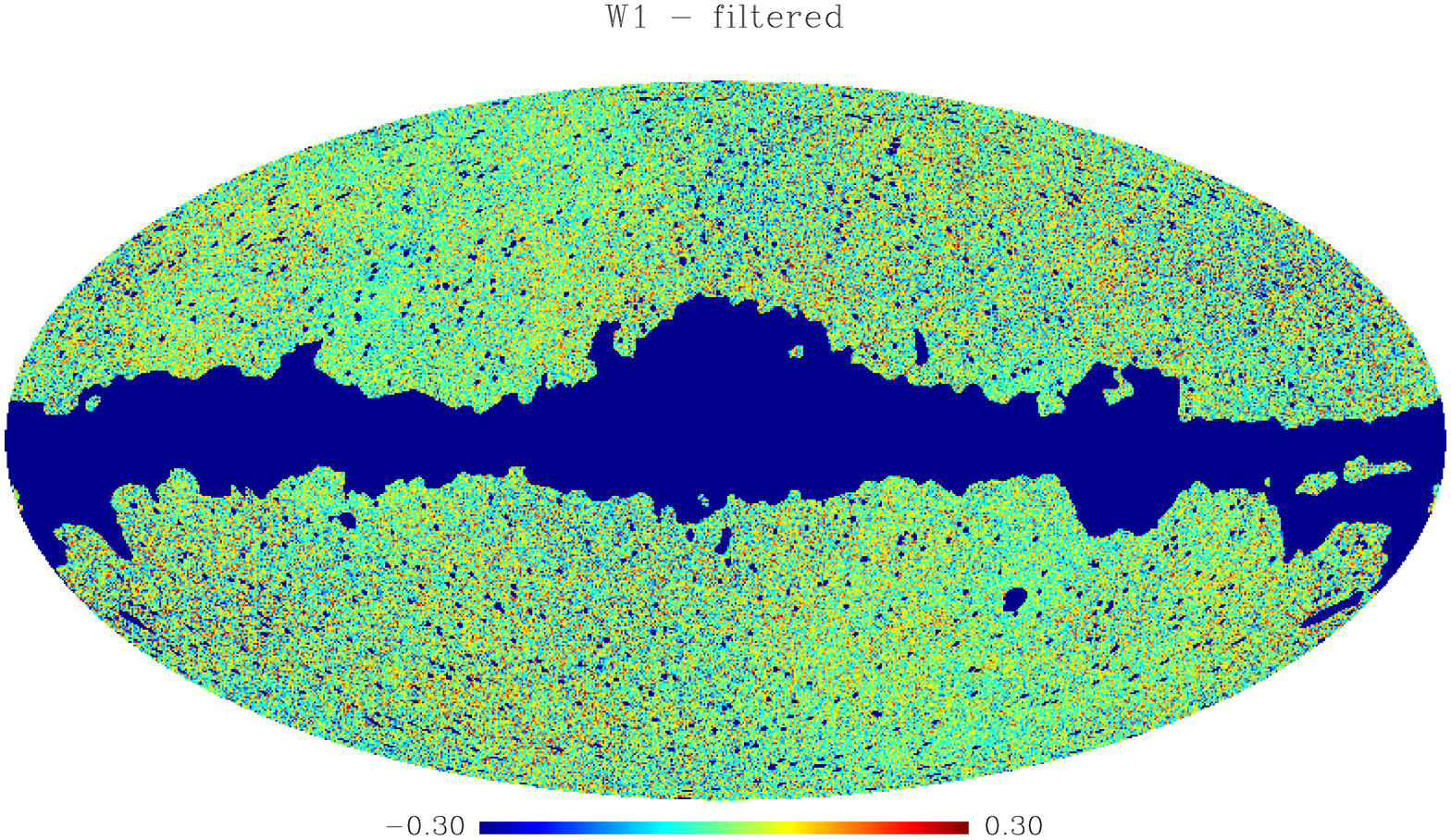} \\
\end{tabular}
\caption{Maps before (left column) and after filtering for
the Q1, V1, W1 channels. The maps are drawn on the same scale. The
KP0 mask is shown with dark blue.\label{fig:maps}}
\end{figure}
\clearpage

The resulting filters are shown for selected channels in Fig.
\ref{fig:filters} for the best-fit $\Lambda$CDM model of the WMAP
team (http://lambda.gsfc.nasa.gov). The filter function is
negative at some of the low $\ell$-multipoles because the true CMB
power spectrum differs from the theoretical input due to cosmic
variance effects. The filter could, in principle, amplify the
noise at low $\ell$, but this effect is very small. We checked
that the filter does not introduce extra variance or correlations.
In any case, larger noise levels in the filtered maps would simply
increase the errors which are measured directly from the same
maps.


Fig. \ref{fig:maps} shows examples of the original and filtered
maps used in our study, and demonstrates that the cosmological CMB
component is removed reliably by the adopted filter.

The SZ components too will be affected by the filter. In
particular, the intrinsic optical depth of the clusters,
determined from X-ray data that have much higher resolution than
WMAP, should be convolved with the filter in any estimate of the
remaining SZ components when using the data from our cluster
catalog.  Because the X-ray pixels are much smaller, the input
$\tau$ should also be convolved with the WMAP beams. Black lines
in Fig.~\ref{fig:filters} show the result product, $B_\ell
F_\ell$, which determines the final effective $\tau$. The
filtering attenuates the $\tau$ profile outside $\sim$10 arcmin.
More power in $\tau$ gets removed in the $\beta$-model, but
filtering will not remove as much power in the more steeply
distributed $\tau$ such as we find in the data.

We demonstrated in a separate study that the extent of the cluster
SZ emission significantly exceeds the one of the X-ray emission
(Atrio-Barandela et al 2008; hereafter AKKE). This is not
surprising because the SZ effect is $\propto n_e$, whereas the
X-ray luminosity $L_X \propto n_e^2$, but, because of the
corresponding decrease in the gas temperature with radius required
by this distribution, it does allow us to integrate down the TSZ
component by selecting pixels within a larger radius, $\alpha
\theta_{\rm X-ray}$ with $\alpha \geq 1$, of the cluster center.
We used $\alpha =[1,2,4,6]$ with a cut at $30^\prime$; at the
largest extent - when we measure the dipole - the angular extents
of clusters become effectively $30^\prime$ across the entire sky.
The reasons for TSZ component washing out sooner than the KSZ one
are that, as measured by us (AKKE) for the same catalog and CMB
data, the cluster X-ray emitting gas is well described by the
density profile expected in the $\Lambda$CDM model (Navarro, Frenk
\& White 1996, hereafter NFW) and the NFW-distributed gas has
X-ray temperature dropping off with radius (e.g. Komatsu \& Seljak
2001); this is discussed in some detail later in the paper. When
extra pixels (not necessarily belonging to the cluster) are added
in the process it would lead to {\it decrease} in the accuracy of
the dipole determination. Our choice of the maximal extent at
$\alpha=6$ is motivated by the measurement that this roughly
corresponds to the maximal extent where the SZ producing gas is
detected on average in the WMAP data (AKKE). Of course, if we were
to increase the total extent further, we should expect that the
dipole component due to KSZ should also start decreasing. We
verified this by computing the CMB dipole from clusters with the
net extent of 1, 2 and 3 degrees. (With this catalog, we cannot go
further since the clusters' overlap starts getting in the way;
e.g. at $3^\circ$ the clusters already occupy $\sim 35\%$ os the
available sky). The decrease in the dipole component is shown in
Fig. \ref{fig:extent} and discussed in detail in Sec.
\ref{results}.

Wiener filtering reduces the TSZ temperature decrement and optical
depth for each cluster. When extending the analysis up to the
largest extent (practically $\simeq 30^\prime$ radius) we find
that the TSZ is diluted by noise and reduced to zero. Since
clusters are not randomly distributed on the sky the TSZ signal
will give rise to a non-trivial dipole signature that, in
principle, may confuse the KSZ dipole. Nevertheless, the dipole
generated by the cross talk with the monopole cannot exceed the
former, i.e. it must be $a_0^{\rm TSZ}>a_{1m}^{\rm TSZ}$, for all
$m$; it is shown below (Table 3) that this component is small. The
following section describes the results of the various simulations
which support this statement.

\section{Error estimation}
\label{errors}

Each of the eight CMB channel maps is processed separately. In the
final maps, we set all pixels to zero that fall outside of both
the cluster areas and the mask and then compute the dipole for
each band using the remove\_dipole procedure in the standard
HEALPix package. Errors are computed from the pixels not
associated with clusters as described below. The results from each
channel are added after weighting with their respective
uncertainties.

We have estimated the errors with two different methods in order
to account for both the effects of the KP0 mask and the intrinsic
distribution of the cluster samples in different redshift bins: 1)
At each $z$-bin we select new random pixels equal to the number of
clusters in each of the eight WMAP channel maps. These new
pseudo-cluster centers are iteratively selected to lie outside the
KP0 mask and away from any of the true cluster pixels. They are
then assigned the cluster radii from the cluster catalog and the
WMAP pixels are selected within these new pseudo-clusters to
compute the new dipole. We then ran 1,000 realizations computing
the errors to within a few percent accuracy. This method accounts
for the effects induced by the geometry of the KP0 mask. 2) In the
second method, we keep the clusters fixed at their celestial
coordinates. The CMB maps for each of the eight channels are then
Fourier transformed and their power spectrum $C_\ell$ computed and
corrected for the fraction of the sky occupied by the KP0 mask. We
use this power spectrum to generate new random phases in the
corresponding $a_{\ell m}$'s, which are then transformed back into
the new CMB sky maps, $T_{\rm new}(\theta,\phi)$. In the new sky
maps we select pixels occupied by the real clusters and compute
the resulting dipole. This method accounts for the effects induced
by the possible leakage from noise and residual CMB due to the
intrinsic distribution of the cluster sample in each $z$-bin.

The two methods give mean zero dipoles with errors that coincide
to within a few percent of each other, which is consistent with
the cluster distribution not confusing the final measurement.

\clearpage
\begin{figure}[h]
\plotone{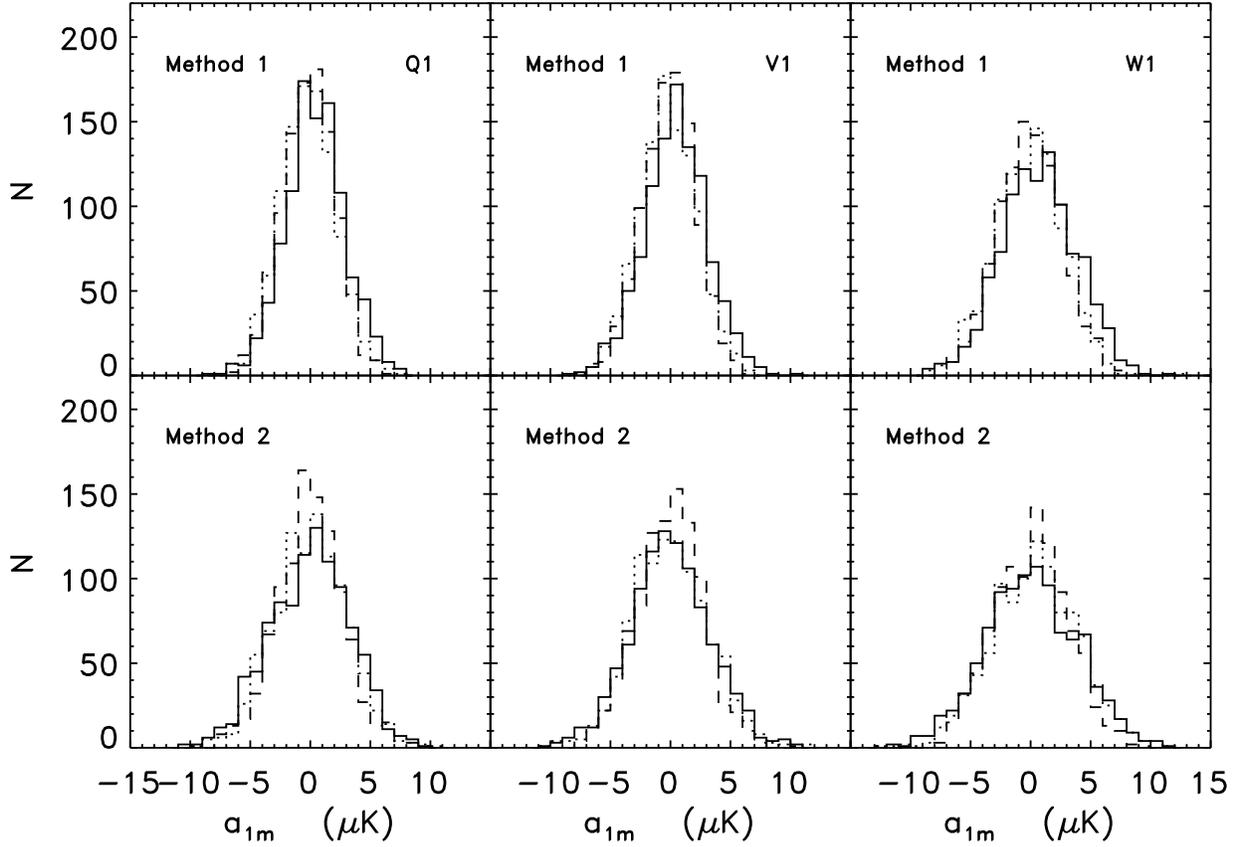}
 \caption{Histograms for
simulations for Q1, V1, W1 channels using as input clusters at
$z\leq 0.05$. Solid lines show the distribution of $a_{1x}$,
dotted for $a_{1y}$ and dashes for $a_{1z}$ from Method 1 (top
panels) and 2. As expected, since the KP0 mask affects most
strongly the $x$-component of the dipole, and least strongly the
$z$-component, the errors on $a_{1x}$ are the largest and on
$a_{1z}$ are the smallest. The largest difference between the
errors from the two methods is for the $x$-component, but even
there the differences are $\lsim 10-15\%$.}
 \label{fig:sims}
\end{figure}
\clearpage

Fig. \ref{fig:sims} shows an example of the distribution of the
dipole components from 1,000 simulations using random pixel
locations in the maps. The figure shows that, as expected, the
distribution of $a_{1m}$ is Gaussian with zero mean, and that the
cosmological CMB component is removed efficiently. The effects of
the CMB mask are such that the largest uncertainty is for the
$a_{1x}$ component of the dipole and the smallest is for $a_{1z}$.
From these simulations we find that the noise terms for $a_{1m}$
integrate down approximately as $\propto N_{\rm cl}^{-1/2}
\alpha^{-1}$, as expected if the CMB component is indeed filtered
out efficiently. Furthermore, we have established that, compared
to the first-year WMAP data, the uncertainties in $a_{1m}$ have
decreased by the expected factor of $\sqrt{3}$. Since the noise
terms are proportional to $t^{-1/2}$, the final 8-year WMAP data
should further improve the measurement.

\clearpage
\begin{figure}[h]
\plotone{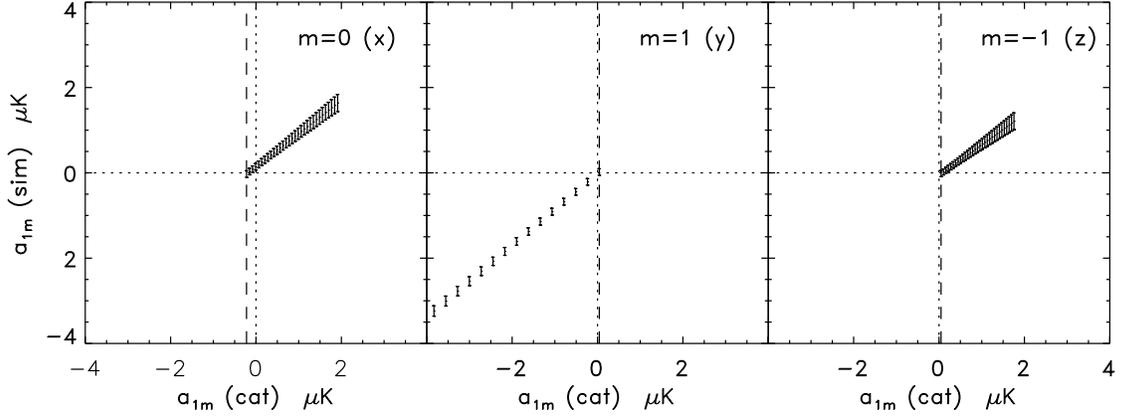}
 \caption{The dipole coefficients for simulated cluster distribution
(random and, on average, isotropic) are compared to that from the
true catalog. (See text for details). Each cluster in each catalog
is given bulk flow of $V_{\rm bulk}$ from 0 to 3,000 km/sec in
increments of 100 km/sec towards the apex of the motion from Table
2. The results from 1,000 simulated catalog realizations were
averaged and their standard deviation is shown in the vertical
axis. Dotted lines mark the zero dipole axis of the panels. Dashed
vertical lines show the dipole due to the modelled TSZ
component.\label{fig:systematics}}
\end{figure}
\clearpage

In order to assess that there is no cross-talk between the
remaining monopole and dipole which may confuse the measured KSZ
dipole, we conducted the following experiment: 1) The TSZ and KSZ
components from the catalog clusters were modelled as described
below in Sec. 6. To exaggerate the effect of the cross-talk from
the TSZ component, the latter was normalized to the {\it maximal}
measured monopole given in Table 3 for the bins where a
statistically significant dipole is detected ($-1.3 \mu$K after
filtering; for comparison Fig. \ref{fig:systematics} shows the
results for the entire catalog, where the measured monopole is
$0\pm 0.2 \mu$K). For the KSZ component each cluster was given a
bulk velocity, $V_{\rm bulk}$, in the direction specified in Table
2, whose amplitude varied from 0 to 3,000 km/sec in 31 increments
of 100 km/sec. The resultant CMB map was then filtered and the CMB
dipole, $a_{1m}({\rm cat})$, over the cluster pixels computed for
each value of $V_{\rm bulk}$. 2) At the second stage we randomized
cluster positions with $(l,b)$ uniformly distributed on celestial
sphere over the {\it full} sky for a net of 1,000 realizations for
each value of $V_{\rm bulk}$ (31,000 in total). This random
catalog keeps the same cluster parameters, but the cluster
distribution now occupies the full sky (there is now no mask) and
on average does not have the same levels of anisotropy as the
original catalog. We then assigned each cluster the same bulk flow
and computed the resultant CMB dipole, $a_{1m}({\rm sim})$, for
each realization. The final $a_{1m}({\rm sim})$  were averaged and
their standard deviation evaluated. Fig. \ref{fig:systematics}
shows the comparison between the two dipoles for each value of
$V_{\rm bulk}$. One can see that there is no significant offset in
the CMB dipole produced by either the mask or the cluster true sky
distribution. The two sets of dipole coefficients are both
linearly proportional to $V_{\rm bulk}$ and to each other; in the
absence of any bulk motion we recover to a good accuracy the small
value of the TSZ dipole. The most noticeable offset is for the
$x$-component of the dipole which is most affected by the mask,
but even here the absolute value of that offset is negligible. In
principle, since the bulk flow motion is fixed in direction and
the cluster distribution is random, one expects the calibration
factor defined below in Sec. \ref{calibration}, $C_{1,100}$ which
translates the dipole in $\mu$K into velocity in km/sec, to be
different from one realization to the next, e.g. in some
realizations certain clusters may be more heavily concentrated in
a plane perpendicular to the bulk flow motion and the measured
$C_{1,100}$ would be smaller. In our case, the mean $C_{1,100}$
differs by $\lsim 10\%$ suggesting that our catalog cluster
distribution is close to the mean cluster distribution in the
simulations. This difference in the overall normalization would
only affect our translation of the dipole in $\mu$K into $V_{\rm
bulk}$ in km/sec.

Finally, we note that the errors computed this way are largely
uncorrelated. For each subsequent $z$-bin we add significantly
more new cluster pixels, but the computed dipole, of course,
includes the clusters in the preceding bins. On the other hand,
the errors are computed from random positions on the maps and
every realization contains, on average, a completely new set of
pixels. There may be some correlations between the various dipole
component errors produced by the mask, but as Figs.
\ref{fig:sims},\ref{fig:systematics} show these correlated
components of the errors are small.

\section{Results}
\label{results}

\subsection{Results by frequency band}

Table 1 shows the measured dipole in the various redshift bins and
shells for each of the three frequency channels (Q, V, W),
combining the numbers from each of the differential assemblies
(DA) with weights obtained from the simulations. One can see that
the dipole amplitude is such that the measurement becomes
statistically significant for $N_{\rm cl}\gsim 300$ for the WMAP
data noise levels. The dipole appears at the negligible monopole
component when computed from the clusters WMAP pixels. By itself
this shows that it cannot originate from the TSZ component.
Nevertheless we also briefly discuss its spectral energy
distribution in as far as it relates to the KSZ origin of the
measured dipole.

The KSZ and TSZ components have different frequency dependence
potentially allowing to distinguish the two origins of the
measured dipole. When CMB photons are scattered by the hot X-ray
gas, the evolution of their occupation number, $n$, is described
by the Kompaneets equation: $ \frac{\partial n}{\partial
y}=\frac{1}{x^2}\frac{\partial}{\partial x}[x^4
\large(\frac{\partial n}{\partial x}+n+n^2\large)]$. Here $x\equiv
h_{\rm P}\nu/k_{\rm B}T_{\rm CMB}$ and $y$ is the comptonization
parameter. In the limit of $y\ll 1$  and for the initially
black-body radiation, $n=1/[\exp(x)-1]$, this equation specifies
the change in the photon spectrum as (e.g. Stebbins 1997):
 \begin{equation}
 \Delta n \simeq y
\frac{x\exp(x)}{[\exp(x)-1]^2}[x {\rm coth}\frac{x}{2}-4]
 \end{equation}
As expected the distortion, $\Delta n$, vanishes at  high
frequency limit ($x\rightarrow \infty$).  The WMAP measurement
data are in thermodynamic temperature units, so the TSZ spectrum
is given by $\Delta T_{\rm TSZ}/T_{\rm CMB} = y G(x)$ with:
 \begin{equation}
G(x) = x {\rm coth} \frac{x}{2}-4
 \label{eq:tsz_spectrum}
 \end{equation}
The expression gives $G(x)$ which is close to $-2$ for low
frequencies, vanishes near 217 GHz, goes positive at higher
frequencies decreasing to zero again at the highest frequencies.
Additionally, there may be non-thermal components and relativistic
corrections (Birkinshaw 1999).

Similarly, the KSZ spectrum can be shown to be given by $\Delta
T_{\rm KSZ}/T_{\rm CMB} = \tau \frac{v}{c} H(x)$ with:
 \begin{equation}
H(x) = 1
 \label{eq:ksz_spectrum}
 \end{equation}
Note that the form of the SZ terms, eqs.(4)-(5), changes if CMB
properties are expressed via the antenna, rather than
thermodynamic, temperature.

The dipole values in Table 1 are flat across the WMAP frequencies,
from 40 to 94 GHz and and are consistent with the spectrum
expected from the KSZ component, although the present data also
give acceptable $\chi^2$ for the TSZ spectrum. Decreasing the
noise by $\sim 2$ expected from the future WMAP measurements may
help distinguish the two components.

\clearpage
\begin{deluxetable}{cc|cccc|ccc}
 \tabletypesize{\scriptsize}
 \tablewidth{0pt}
 \tablecaption{Results from Q, V, W filtered maps.}
  \tablehead{
 \colhead{} & \colhead{} & \multicolumn{4}{c}{Multipoles} &
 \multicolumn{3}{c}{Shells: $0.05 < z \leq $ \& $0.12 < z \leq 0.3$ }\\
 \cline{3-6}
 \cline{7-9}
 \colhead{$z\leq$} &
 \colhead{Band} &
 \colhead{$\langle \Delta T\rangle$} &
 \colhead{$a_{1,{\rm x}}$} &
 \colhead{$a_{1,{\rm y}}$} &
 \colhead{$a_{1,{\rm z}}$} &
 \colhead{$a_{1,{\rm x}}$} &
 \colhead{$a_{1,{\rm y}}$} &
 \colhead{$a_{1,{\rm z}}$}
}
 \startdata
  & & $\mu$K & $\mu$K & $\mu$K & $\mu$K & $\mu$K & $\mu$K & $\mu$K
  \\
  \hline
0.05 & Q & $-0.1 \pm 0.9$ & $-1.0 \pm 1.7$ & $-3.6 \pm 1.6$ & $0.0 \pm 1.5$ & \nodata & \nodata & \nodata \\
\nodata & V & $0.8\pm 1.0$ & $-1.3 \pm 1.8$ & $-2.9 \pm 1.6$ & $-0.2\pm 1.5$ & \nodata & \nodata & \nodata \\
\nodata & W & $-0.4\pm 0.9$ & $-0.3\pm 1.7$ & $-3.2 \pm 1.5$ & $0.1\pm 1.4$ & \nodata & \nodata & \nodata \\
0.06 & Q & $-1.0\pm 0.8$ & $-0.4 \pm 1.5$ & $-2.6 \pm 1.3$ & $-0.7 \pm 1.3$ & \nodata & \nodata & \nodata \\
\nodata & V & $-0.0 \pm 0.8$ & $-1.1 \pm 1.5$ & $-2.2 \pm 1.4$ & $-0.9 \pm 1.3$ & \nodata & \nodata & \nodata \\
\nodata & W & $-1.6 \pm 0.7$ & $-1.5 \pm 1.4$ & $-2.2 \pm 1.3$ & $0.1 \pm 1.2$ & \nodata & \nodata & \nodata \\
0.08 & Q & $-1.1 \pm 0.7$ & $ 1.8 \pm 1.2$ & $-1.4\pm 1.1$ & $-1.6 \pm 1.0$ & \nodata & \nodata & \nodata \\
\nodata & V & $-0.5 \pm 0.7$  & $1.5 \pm 1.2$ & $-1.6\pm 1.1$ & $-1.2 \pm 1.0$ & \nodata & \nodata & \nodata \\
\nodata & W & $-2.1 \pm0.6$ & $0.5 \pm 1.1$ & $-1.5 \pm 1.0$ & $-0.5 \pm 0.9$ & \nodata & \nodata & \nodata \\

0.12 & Q & $-0.8 \pm 0.5$ & $1.4 \pm 1.0$ & $-1.8 \pm0.9$ & $-0.6 \pm 0.8$ & $2.8 \pm 1.2$ & $-0.9 \pm 1.1$ & $-0.9 \pm 1.0$ \\
\nodata & V & $-0.3\pm 0.5$ & $1.7 \pm 1.0$ & $-2.2 \pm 0.9$ & $-0.4 \pm 0.9$ & $3.7 \pm 1.2$ & $-1.8 \pm 1.1$ & $-0.4 \pm 1.0$ \\
\nodata & W & $-1.1 \pm 0.5$ & $1.4\pm 0.9$ & $-2.5 \pm 0.8$ & $-0.2 \pm 0.8$ & $2.7 \pm 1.1$ & $-2.2 \pm 1.0$ & $-0.5 \pm 0.9$ \\

0.16 & Q & $-0.1 \pm 0.5$ & $0.8 \pm 0.9$ & $-2.6 \pm 0.8$ & $-0.1 \pm 0.8$ & $1.6 \pm 1.0$ & $-2.2 \pm 0.9$ & $-0.2 \pm 0.9$ \\
\nodata & V & $0.4 \pm 0.5$ & $1.1 \pm 0.9$ & $-2.6 \pm 0.8$ & $0.4 \pm 0.8$ & $2.3 \pm 1.1$ & $-2.6 \pm 0.9$ & $0.6 \pm 0.9$ \\
\nodata & W & $-0.4 \pm 0.4$ & $0.1 \pm 0.9$ & $-3.5 \pm 0.8$ & $0.2 \pm 0.7$ & $0.5 \pm 1.0$ & $-3.7 \pm 0.9$ & $0.1 \pm 0.8$ \\

0.20 & Q & $-0.0\pm 0.5$ & $1.0 \pm 0.9$ & $-2.9 \pm 0.8$ & $0.3 \pm 0.7$ & $1.8 \pm 1.0$ & $-2.7 \pm 0.9$ & $0.5 \pm 0.8$ \\
\nodata & V & $0.5 \pm 0.5$ & $1.1 \pm 0.9$ & $-2.8 \pm 0.8$ & $0.7 \pm 0.7$ & $2.2 \pm 1.0$ & $-2.9 \pm 0.9$ & $1.0 \pm 0.8$ \\
\nodata & W & $-0.2 \pm 0.4$ & $0.2 \pm 0.8$ & $-4.1 \pm 0.7$ & $0.6 \pm 0.7$ & $0.6 \pm 0.9$ & $-4.4 \pm 0.9$ & $0.6 \pm 0.8$ \\

0.30 & Q & $-0.1 \pm 0.4$ & $0.9 \pm 0.8$ & $-2.2 \pm 0.7$ & $0.4 \pm 0.7$ & $1.6 \pm 0.9$ & $-1.9 \pm 0.8$ & $0.5 \pm 0.8$ \\
\nodata & V & $0.4 \pm 0.4$ & $1.2 \pm 0.9$ & $-2.2 \pm 0.8$ & $0.7 \pm 0.7$ & $2.2 \pm 0.9$ & $-2.2 \pm 0.8$ & $1.0 \pm 0.8$ \\
\nodata & W & $-0.3 \pm 0.4$ & $-0.2 \pm 0.8$ & $-3.5 \pm 0.7$ & $0.5 \pm 0.6$ & $0.1 \pm 0.9$ & $-3.7 \pm 0.8$ & $0.5 \pm 0.7$ \\
 \hline
 0.12--0.3 & Q & \nodata & \nodata & \nodata & \nodata & $0.8 \pm
 1.4$ & $-2.9 \pm 1.3$ & $2.3 \pm 1.2$ \\
 \nodata & V & \nodata & \nodata & \nodata & \nodata & $0.9 \pm
 1.5$ & $-2.2 \pm 1.3$ & $3.1 \pm 1.2$ \\
 \nodata & W & \nodata & \nodata & \nodata & \nodata & $-2.2 \pm
 1.4$ & $-5.2 \pm 1.3$ & $1.9 \pm 1.2$ \\
\enddata
\tablecomments{Intermediate results are shown for each of the WMAP
bands in the the redshift bins specified in the first column.
Columns 3-6 give the numbers for the standard cluster
configuration used in the paper. The last three columns show the
dipole in the shell configuration excluding the clusters at $z\leq
0.05$. In the latter case we restricted the runs to when we are
left with at least 300 clusters in the shell in order to get
statistically meaningful results.}
\end{deluxetable}
\clearpage

As a further consistency check and to estimate how much of the
signal is contributed by the farthest clusters, we have also
computed the numbers in a shell configuration excluding clusters
with $z\leq 0.05$ and for the 274 clusters with $0.12 \leq z \leq
0.3$. Interpretation of such numbers can be cumbersome because of
the complicated window involved, but nevertheless they can provide
a useful diagnostic of the consistency of the results and the
contribution to the dipole by the farthest clusters. Our results
show that we start getting statistically meaningful results with
at least $\sim 300$ clusters, so the runs were done for the bins
where the outer $z$ exceeded 0.012. The dipole coefficients for
each band are shown in the last three columns of Table 1. They are
overall consistent with the main results and provide further
support that the dipole is generated by cluster motions on the
largest scales.

\clearpage
\begin{figure}[h]
\plotone{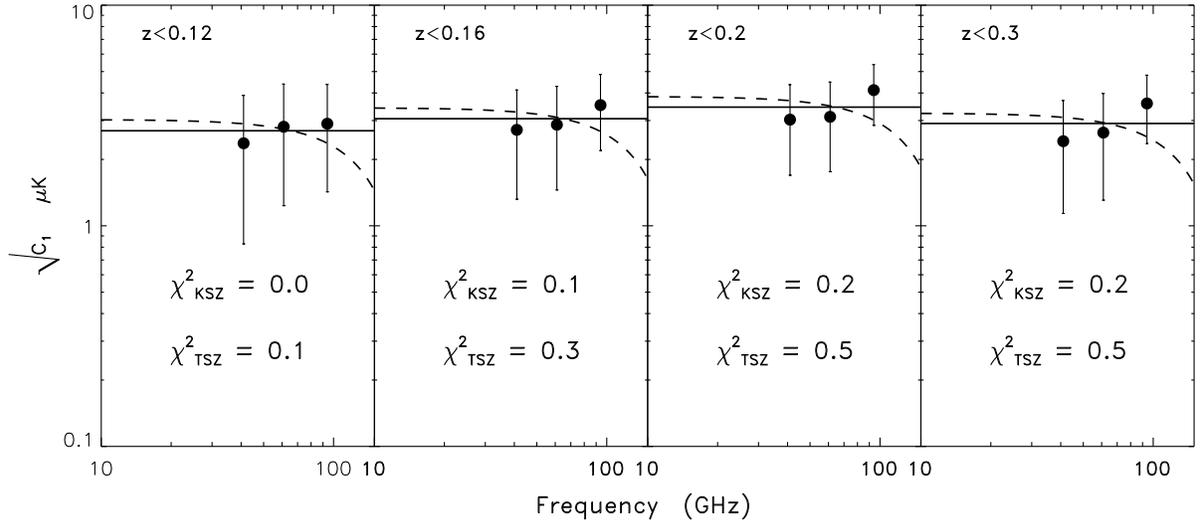}
 \caption{Spectral energy distribution of the measured dipole amplitude vs the frequency of
 each of the WMAP Q, V, W bands. The measured amplitudes are shown with circles and 1-$\sigma$
 uncertainties. Solid lines show the spectrum of any KSZ component, given by eq.
\ref{eq:ksz_spectrum} obtained by minimizing the corresponding
$\chi^2$; dashed lines show the same for the TSZ component given
by eq. \ref{eq:tsz_spectrum}. The corresponding $\chi^2$ per two
degrees of freedom are also shown in the panels.\label{fig:sed}}
\end{figure}
\clearpage

Fig. \ref{fig:sed} plots the dipole amplitude for four farthest
redshift bins vs the frequency of each channel juxtaposed against
the TSZ energy spectrum normalized to the measured dipole at 40
GHz. The TSZ spectrum (eq. \ref{eq:tsz_spectrum} below) would
predict a smaller dipole value in the W band. On the other hand,
the spectrum of the dipole arising from the KSZ should be flat
across the frequencies consistent with the plotted numbers (as
mentioned above and shown in the figure the TSZ spectrum also
gives acceptable $\chi^2$ given the noise in the present WMAP
data).

\subsection{Results averaged over all frequency channels.}

Table 2 shows the results after weight-averaging over all of the
eight DA's. The table also gives additional information on the
cluster samples used in each measurement. In order to assess the
potential impact of cooling flows on the results, we have also
made the computations omitting cluster central pixels in WMAP
data. The results were essentially unchanged compared to those
presented in the table. There is a clear statistically-significant
dipole at the level of $\sim 2-3 \mu$K once we reach $\sim 300$
clusters and the aperture ($\simeq 30^\prime$) encompassing most
of the hot gas producing the SZ effect. {\it The dipole remains as
the monopole representing the mean TSZ component from hot gas
within the selected aperture vanishes}.

\clearpage
\begin{deluxetable}{c c c c c c c c c c c c c}
\tablewidth{0pt} \tabletypesize{\scriptsize} \rotate
\tablecaption{Cluster and map parameters with results from
averaging over all channels.
 \label{table} }
 \tablehead{
 \colhead{(1)} & \colhead{(2)} & \colhead{(3)} & \colhead{(4)} &
 \colhead{(5)} & \colhead{(6)} & \colhead{(7)} &
 \colhead{(8)} &
 \colhead{(9)} & \colhead{(10)} & \colhead{(11)} & \multicolumn{2}{c}{(12)} \\
 \colhead{$z\leq$} &
 \colhead{$\langle z\rangle$}  & \colhead{$z_{\rm median}$} & \colhead{$N_{\rm cl}$} &
\colhead{$N_{\rm pixels}$} & \colhead{$\langle T \rangle$} &
\colhead{$a_{1,x}$} & \colhead{$a_{1,y}$} & \colhead{$a_{1,z}$} &
\colhead{$\sqrt{C_1}$} & \colhead{(l,b)} &
 \multicolumn{2}{c}{$\sqrt{C_{1,100}}$: $\mu$K per 100 km/sec} \\
 \cline{12-13}
 \colhead{}& \colhead{} & \colhead{} & \colhead{} &  \colhead{} &
\colhead{$\mu$K} &
 \colhead{$\mu$K} & \colhead{$\mu$K} & \colhead{$\mu$K} & \colhead{$\mu$K} & \colhead{deg}
 & \colhead{(a)} & \colhead{(b)} }
 \startdata
  0.02  & 0.016 & 0.016 & 17 & 941 & $-2.6\pm 1.5$ & $-4.4 \pm 3.2$ & $2.0 \pm 2.9$ & $7.5 \pm
  2.5$ & $8.9 \pm 5.0$ & n/a & 0.18 (0.70) & 0.20 (0.84) \\
  0.025 & 0.018 & 0.019 &  27 & 1,497 & $-5.2 \pm 1.2$ & $-5.5 \pm 2.4$ & $-2.9 \pm 2.1$ & $0.1 \pm
  2.0$ & $6.2 \pm 3.8$ & n/a & 0.18 (0.70)  & 0.20 (0.78) \\
  0.03  & 0.022 & 0.023 & 43 & 2,417 & $-5.9 \pm 1.0$ & $0.3 \pm 1.9$ & $2.9 \pm 1.6$ & $0.5 \pm
  1.6$ & $3.0 \pm 2.9$ & n/a & 0.18 (0.73) & 0.20 (0.82) \\
  0.04  & 0.029 & 0.030 & 86 & 4,872 & $0.5 \pm 0.7$ & $-1.6 \pm 1.3$ & $0.8 \pm 1.1$ & $-2.6 \pm
  1.1$ & $3.1 \pm 2.0$ & n/a & 0.20 (0.73) & 0.24 (0.89) \\
  0.05  & 0.035 & 0.036 & 135 & 7,575 & $0.1 \pm 0.5$ & $-0.8 \pm 1.0$ & $-3.3 \pm 0.9$ & $-0.0 \pm
  0.9$ & $3.4 \pm 1.6$ & $(256,-0)\pm 24$ & 0.22 (0.76) & 0.24 (0.82) \\
  0.06 & 0.041 & 0.042 & 188 & 10,474 & $-1.0 \pm 0.4$ & $-1.0 \pm 0.9$ & $-2.4 \pm 0.8$ & $-0.5 \pm
  0.7$ & $2.6\pm 1.4$ & $(247,-10)\pm 26$ & 0.22 (0.80) & 0.22 (0.79) \\
  0.08 & 0.051 & 0.053 & 292 & 16,064 & $-1.3 \pm 0.4$ & $1.3 \pm 0.7$ & $-1.5 \pm 0.6$ & $-1.1 \pm
  0.6$ & $2.2 \pm 1.1$ & $(310,-29)\pm 24$ & 0.24 (0.76) & 0.26 (0.83) \\
  0.12 & 0.067 & 0.067 & 444 & 24,189 & $-0.7 \pm 0.3$ & $1.5 \pm 0.6$ & $-2.2 \pm 0.5$ & $-0.4 \pm
  0.5$ & $2.7 \pm 0.9$ & $(305, -9)\pm 17$ & 0.26 (0.79) & 0.28 (0.88) \\
  0.16 & 0.080 & 0.076 & 541 & 29,127 & $-0.1 \pm 0.3$ & $0.7 \pm 0.5$ & $-2.9 \pm 0.5$ & $0.1 \pm
  0.4$ & $3.0 \pm 0.8$ & $(283, 3)\pm 13$ & 0.25 (0.75) & 0.27 (0.83) \\
  0.20 & 0.090 & 0.082 & 603 & 32,146 & $0.1 \pm 0.3$ & $0.7 \pm 0.5$ & $-3.3 \pm 0.4$ & $0.5 \pm
  0.4$ & $3.4 \pm 0.8$ & $(282, 9)\pm 11$ & 0.28 (0.84) & 0.29 (0.90) \\
  All $z$ & 0.106 & 0.089 & 674 & 35,409 & $0.0 \pm 0.2$ & $0.6 \pm 0.5$ & $-2.7 \pm 0.4$ & $0.6 \pm
  0.4$ & $2.8 \pm 0.7$ & $(283,11)\pm 14 $ & 0.29 (0.965) & 0.32 (1.01) \\
  \hline
  0.05-0.3 & 0.12 & 0.11 & 540 & 29,896 & $-0.1 \pm 0.3$ & $1.2
  \pm 0.5$ & $-2.6 \pm 0.5$ & $0.7 \pm 0.4$ & $2.9 \pm 0.8$ &
  $(295,14)\pm 13$
  & 0.31 (0.84) & 0.33 (0.92) \\
  0.12-0.3 & 0.18 & 0.17 & 230 & 11,920 & $1.7 \pm 0.4$ & $-0.2
  \pm 0.8$ & $-3.5 \pm 0.7$ & $2.4 \pm 0.7$ & $4.2 \pm 1.3$ &
  $(267,34)\pm 15$
  & 0.36 (0.89) & 0.40 (1.0) \\
 \enddata
\tablecomments{Results are shown for the KP0 mask only with the SZ
cluster extent taken to be $\min[6\theta_{\rm X-ray},30^\prime]$.
All uncertainties correspond to 1$\sigma$ from Method 1 in Sec.
\ref{errors}; the errors are from 1,000 realizations, so the error
uncertainty is $\simeq 4\%$. Method 2 gives identical errors
within $\lsim 10\%$. E.g.: at $z\leq 0.05$, where we first recover
a statistically significant dipole, the errors from Method 2 are
$(1.16, 1.09, 0.94)\mu$K for the $(x,y,z)$ dipole; at $z\leq 0.3$
they become $(0.62,0.56,0.46)\mu$K. By the time the results are
rounded to one significant digit in the table the two sets have
little difference and for brevity only one set of errors is shown.
Of course, the monopole errors are the same for the two methods.
The columns are: (1)-(3) the upper, mean and median redshift of
the cluster bins. (4),(5) The number of clusters and the number of
pixels used in evaluating the dipole in each redshift bin. (6) The
mean CMB temperature evaluated over the cluster pixels in each
bin. (7)-(10) Three dipole components, $a_{1m}$, and the dipole
amplitude, $\sqrt{C_1}$, evaluated over the cluster pixels in each
bin. (11) Direction and its uncertainty associated with the CMB
dipole shown for the redshift bins where there is a statistically
significant (at least $2\sigma$) measurement of $\sqrt{C_1}$. (12)
The total dipole amplitude for $V_{\rm bulk}$=100 km/sec for
filtered and unfiltered (in parentheses) maps determined using
$r_c$ and $n_e$ values for each cluster obtained via (a) our
best-fit $\beta$-model to the RASS data and (b) from the empirical
relationship as described below The top 11 rows correspond to
sphere configurations; the last two rows correspond to clusters in
shells. Of the latter, the last shell has median dipole of $\simeq
0.18$ showing that the measured dipole is produced by the
outermost clusters at median depth of $\gsim 600h^{-1}$Mpc.
Previously claimed peculiar flows had directions: i) CMB dipole is
in the direction of
$(l,b)=(264.26^\circ\pm0.33^\circ,48.22^\circ\pm0.13^\circ)$ and
after correction for the Local Group motion becomes towards
$(l,b)=(276^\circ \pm 3^\circ, 30^\circ\pm 3^\circ)$ (see
\cite{strauss-willick} and references therein); ii) the Great
Attractor motion based on the Fundamental Plane distance indicator
\cite{7s-di,djorgovski} is towards $(l,b)=(307,9)^\circ$
\cite{7s-motion}; iii) using brightest cluster galaxies as
distance indicators by \cite{lauer-postman} gave motion toward
$(l,b)=(220,-28)^\circ$ with uncertainty of $\pm27^\circ$; iv)
Analysis of a sample of spiral galaxies using the Tully-Fisher
relation as distance indicator by \cite{willick99} suggested
motion to $(l,b)=(342,52)^\circ$ with $\pm 23^\circ$ uncertainty;
v) ref. \cite{hudson} use early galaxy sample for 56 clusters and
find motion to $(l,b)=(260 \pm 15,-1\pm 12)^\circ$.}
\end{deluxetable}
\clearpage

The direction of the dipole and its uncertainty in Table 2 were
computed as follows: each dipole component is assumed
Gaussian-distributed, with the given mean and errors. At each $z$
we generate $10^4$ dipoles from a normal distribution with the
standard deviation equal to each component error bar and compute
the angle of these dipoles with respect to the direction of the
mean dipole. For small angles, this angle follows a $\chi^2$
distribution with 3 degrees of freedom; the uncertainty in the
table corresponds to the 68 \% confidence contour of this
distribution. The directions from previous measurements of
peculiar flows based on galaxy distance indicators and those of
the acceleration dipoles of the various cluster studies are
summarized in the note to Table 2. The direction of the bulk flow
deduced here is $\sim20^{\circ}$ from the ``global CMB dipole"
direction, with a 1-$\sigma$ error of $\sim 10^\circ$-$25^\circ$
over the range of $z$ probed in this study, and does not vary
significantly within the range covered by our data.

\clearpage
\begin{figure}[h]
\plotone{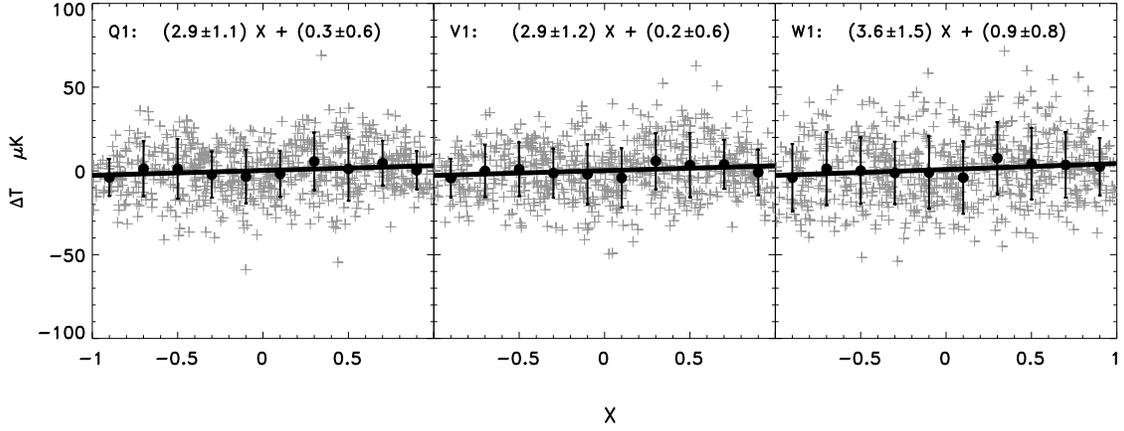}
 \caption{CMB temperature (light shaded plus signs) for each of 674 clusters out to $z\leq 0.3$ is plotted
vs $X$, the cosine of the angle between the dipole apex and each
cluster. The plots are shown for one DA channel at each frequency.
The linear fit to the data is shown with thick solid line; its
parameters and their uncertainties are displayed at the top of
each panel. The uncertainties in the displayed fits were computed
using uniform weighting. Filled circles with errors show the mean
and standard deviation over all clusters binned in ten equally
spaced bins in $X$. The correlation coefficient of the binned data
shown with circles, $r={\rm cor}(X,\Delta T)$, is 0.5 in Q1, V1
bands and 0.6 in W1 band. For the unbinned data the correlation
coefficient is $\simeq 0.1$ in each of the channels, whereas the
random uncorrelated data would give $r=0$ to within
$1/\sqrt{N_{\rm cl}}=0.038$; this is another way of saying that we
detect the dipole at $\sim 2.5\sigma$ level at each channel.}

 \label{fig:2apex}
\end{figure}
\clearpage

The reality of the measured dipole can also be seen in from the
following: In Fig. \ref{fig:2apex} we present the measured signal
of the entire cluster sample ($z\leq 0.3$) plotted against $X$,
the cosine of the angle between the detected dipole and the
cluster itself for three channels at three different frequencies
(Q1, V1, W1). For each cluster the CMB temperature was averaged
over the cluster pixels out to $\min[6\theta_{\rm X},30^\prime]$.
Results from linear fits (thick solid lines) to the data and their
uncertainties are displayed in each panel. As expected there is a
statistically significant dipole component in the cluster CMB
temperatures. In each of the eight channels the significance is $>
2 \sigma$ leading to the overall result in the main text. The
signal is consistent with the spectrum expected from the KSZ
component.

\section{TSZ monopole vs KSZ dipole and related issues}
\label{tsz}

\begin{figure}[h]
\plotone{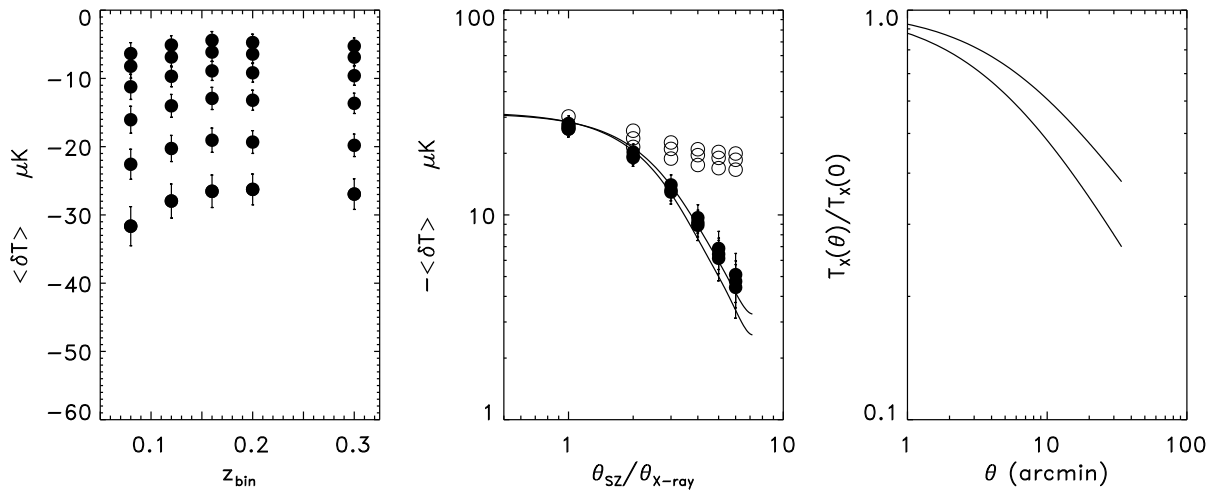}
 \caption[]{\small{{\bf Left}: The mean CMB temperature decrement
 averaged over the Q, V, W channels. The results are for unfiltered
maps with 0.5$^\circ$ cut in cluster extent shown for the outer
$z$-bins for progressively increasing $\alpha=\theta_{\rm
SZ}/\theta_{\rm X-ray}$. Filled circles from bottom to top
correspond to $\alpha=1,2,3,4,5,6$. {\bf Middle}: Solid circles
show the mean TSZ decrement profile in the unfiltered CMB data vs
$\alpha$ for three farthest $z$-bins. Open circles correspond to
the isothermal $\beta=2/3$ model evaluated as described in Sec. 6.
The two solid lines correspond to the NFW profile with
concentration parameter $c=6,10$ normalized to the mean cluster
parameters (see AKKE for details). The measured decrease in the
filtered TSZ monopole is shown in Fig. 1 of Kashlinsky et al
(2008). {\bf Right}: the X-ray temperature profile in units of the
temperature at the center for the NFW profiles shown in the middle
panel. The angular scale $\theta$ in arcmin corresponds to the
average X-ray extent of our cluster sample. I.e. the NFW profile
corresponds to a single cluster of virial radius $2h^{-1}$Mpc
located at an angular distance $d_A=250 h^{-1}$Mpc. }}
\label{fig:tsz}
\end{figure}

We demonstrate in AKKE that our cluster catalog applied to the
{\it unfiltered} CMB data indicates that the gas in X-ray clusters
is well described by the Navarro-Frenk-White (1996, NFW) density
profile theoretically expected from the non-linear evolution of
the concordance $\Lambda$CDM model. In addition to using
unfiltered maps, the analysis of that paper was done without
imposing the 30$^\prime$ cut on the maximal cluster extent,
defined a different effective cluster angular scale and the table
there shows the monopole averaged over all the DA's with very
different angular resolution diluting the underlying true TSZ
signal. Hence, here we revisit their conclusions for the dataset
used throughout this measurement. In the left panel of Fig.
\ref{fig:tsz} we show the mean TSZ decrement at the cluster
positions evaluated from the WMAP maps for the various total
cluster extent limits described in Sec. 1 (as discussed, the
maximal extent here is truncated at 30$^\prime$). The errors are
standard deviations of the CMB temperature evaluated with 1,000
random realizations of pseudo-clusters over the CMB map pixels
outside the mask and away from the catalog clusters. The mean
temperature decrement from each of the eight DA's were
weight-averaged with their corresponding uncertainties to give the
final $\langle \delta T \rangle$ shown in the figure. The strong
decrease in the mean TSZ decrement with the increasing angular
size is apparent from the figure.
\clearpage
\clearpage

The middle panel of the figure shows the mean CMB temperature
profile of the TSZ decrement in the unfiltered maps for three
outer redshift bins. (The decrease of the filtered TSZ decrement
profile is shown in Fig. 1 of Kashlinsky et al, 2008). The
expectation from the isothermal $\beta$-model for these bins was
evaluated as described in Sec. 6 and is shown with the open
circles. It fits well the data at the cluster inner parts, but
deviates strongly from the measurements at larger radii. The fits
from the NFW profiles using a method similar to Komatsu \& Seljak
(2001) are shown with solid lines for two concentration parameters
(see AKKE for details). These profiles provide a good fit to the
data.

It is important to emphasize in this context that the gas with the
NFW profile which is in hydrostatic equilibrium with the cluster
gravitational field must have the X-ray temperature decreasing
with radius (Komatsu \& Seljak 2001). This is confirmed by
numerical simulations of the cluster formation within the
$\Lambda$CDM model (Borgani et al 2004) as well as by the
available observations of a few nearby clusters (Pratt et al
2007). The latter cannot yet probe the $T_{\rm X}$ profile all the
way to the virial radius, but do show a decrease by a factor of
$\sim 2$ out to about half of it (see e.g. Fig. 5 of Pratt et al
2007). In the NFW profile the gas density profile in the outer
parts goes as $n_e \propto r^{-3}$ with the polytropic index which
is approximately constant for all clusters at $\gamma \simeq 1.2$
(Komatsu \& Seljak 2001). Thus the X-ray temperature must drop at
least as $T_{\rm X} \propto r^{-0.6}$ at the outer parts and for
larger values of $\gamma$ the drop will be correspondingly more
rapid. The temperature profile implied by the NFW density profile
normalized to the data in the middle panel is shown in the right
panel of Fig. \ref{fig:tsz}.

\clearpage
\begin{deluxetable}{cccccc}
 \tabletypesize{\scriptsize}
 \tablewidth{0pt}
 \tablecaption{TSZ monopole vs KSZ dipole contributions from rings.} \tablehead{
 \colhead{Ring} & \colhead{$N_{\rm pixels}$} &
 \multicolumn{1}{c}{Monopole} & \multicolumn{3}{c}{Dipole components (filtered)} \\
 \cline{3-3}
 \cline{4-6}
\colhead{} &
\\

 \colhead{ } & \colhead{ } &
 \colhead{(unfiltered)} &
 \colhead{$a_{1,{\rm x}}$} &
 \colhead{$a_{1,{\rm y}}$} &
 \colhead{$a_{1,{\rm z}}$}

}
  \startdata
  & & $\mu$K & $\mu$K & $\mu$K & $\mu$K \\
  \hline
$0^\prime-5^\prime$ & 1,183 & $-24.5\pm 9.2$ & $3.5 \pm 4.4$ & $-0.9 \pm 3.7$ & $-6.2 \pm 3.5$ \\
$5^\prime-10^\prime$ & 3,283 &  $-18.0 \pm 5.5$ & $1.2 \pm 2.6$ & $-4.4 \pm 2.2$ & $-5.2 \pm 2.1$ \\
$10^\prime-15^\prime$ & 5,546 & $-12.6 \pm 4.3$ & $2.2 \pm 2.0$ & $-5.2 \pm 1.7$ & $2.9 \pm 1.6$ \\
$15^\prime-20^\prime$ & 7,673 & $-6.8 \pm 3.6$ & $0.6 \pm 1.7$ & $-4.8 \pm 1.5$ & $2.0 \pm 1.4$ \\
$20^\prime-25^\prime$ & 9,744 & $-6.0 \pm 3.2$ & $-0.3 \pm 1.5$ & $-2.8 \pm 1.3$ & $0.5 \pm 1.2$ \\
$25^\prime-30^\prime$ & 11,845 & $-5.8 \pm 2.9$ & $0.9 \pm 1.4$ & $-1.0 \pm 1.2$ & $-0.3 \pm 1.1$ \\
$30^\prime-45^\prime$ & 47,064 & $-4.6 \pm 1.5$ & $2.7 \pm 0.7$ & $-2.0 \pm 0.6$ & $1.4 \pm 0.6$  \\
$45^\prime-60^\prime$ & 63,987 & $-4.3 \pm 1.3$ & $0.5 \pm 0.6$  & $-0.7 \pm 0.5$ &  $-0.9 \pm 0.5$ \\
\enddata
\tablecomments{Differential contributions to the TSZ monopole (for
original maps) and to the KSZ dipole using the lowest resolution
W-band (FWHM $\simeq 0.2^\circ$) and the entire cluster catalog
with $z\leq 0.3$ and KP0 mask; the ring width is smaller than the
resolution of the Q, V WMAP bands. Negative monopole values in the
original maps are expected from the TSZ component. The
measurements show the existence of the hot intracluster gas out
$\gsim 30^\prime$ confirming that the dipole is traced by the KSZ
component from the cluster gas.}
\end{deluxetable}
\clearpage

The implications of the above are that in the outer parts of
clusters the TSZ monopole component must decrease faster than the
KSZ dipole as we increase the aperture to probe the cluster outer
regions. This is what we observe in the data and allowed us to
isolate the KSZ dipole component as the TSZ monopole vanishes. The
reason we present the results out to $\min[6\theta_{\rm
X-ray},30^\prime]$ is that this is roughly the scale where we
still detect statistically significant TSZ signal in the
unfiltered data (AKKE and Table 3 below).

Table 3 shows the differential distribution of the TSZ and dipole
components in rings of the specified radius and width around the
clusters in our catalog. The data from the W-band were selected
for the table because this channel has the finest angular
resolution making it the most adequate to probe the differential
contribution to the final signal. The table clearly shows that in
the unfiltered data the X-ray emitting gas producing the TSZ
signal exists out to at least $25^\prime-30^\prime$, the effective
final radius of our cluster catalog. The measurements in the table
confirm explicitly that, due to the X-ray temperature decrease, in
the filtered maps the dipole KSZ component can be isolated as the
TSZ monopole vanishes. Mathematically, discounting the additional
$T_{\rm X}$ factor with $\gamma\simeq 1.2-1.3$ in the TSZ terms
makes the KSZ term for the NFW-like clusters lie close to the TSZ
profile of the isothermal $\beta\sim2/3$ model and, hence, its KSZ
decrease with increasing aperture radius should roughly mimic the
open circles in the middle panel of Fig. \ref{fig:tsz}. Of course,
the true cluster properties, such as the electron density and
X-ray temperature profiles, can be mathematically constrained (and
perhaps even recovered) from the measurements of both the KSZ
dipole and TSZ monopole profiles; this, however, lies outside the
scope of this investigation. \clearpage
\begin{figure}[h]
\plotone{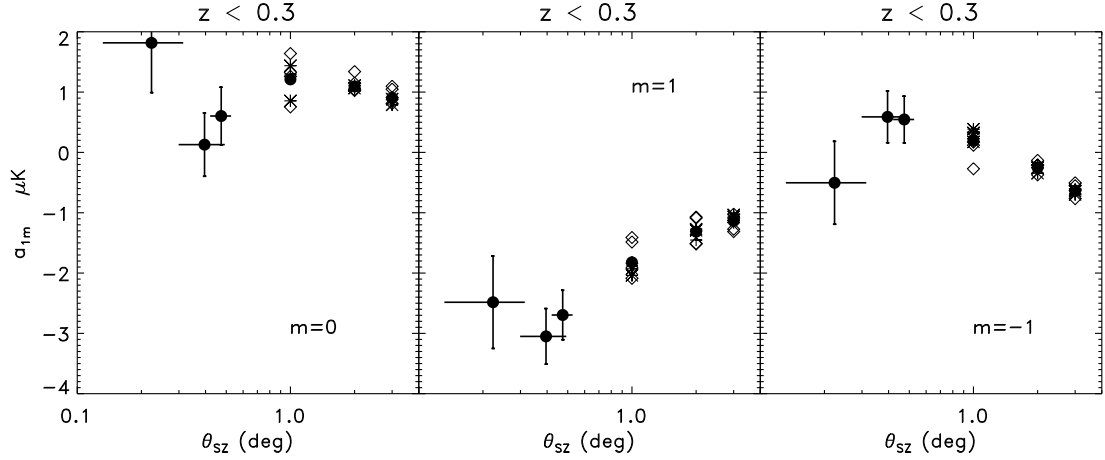}
 \caption{Shows the dipole variation with increasing cluster extent.
The cluster electron density profile is such that within the
statistical uncertainties the dipole does not change significantly
out to the cluster virial radius, where we still detect
statistically significant TSZ component as shown in Fig.
\ref{fig:tsz}. At larger radii it starts decreasing as expected,
although if there was some hot gas around the clusters the
decrease would be slower. The last three sets of points show the
dipole as all clusters are set to extend to 1, 2 and 3 degree
radii; the first three correspond to $\alpha = 2,4,6$ and the
maximal extent set at 0.5 degrees as shown in Fig. 1 of Kashlinsky
et al (2008). Pluses correspond to the two Q channel DA's,
asterisks to the two V channel DA's and diamonds to the four W
channel DA's. Filled circle shows the mean over all eight DA's.
The horizontal error bar shows the standard deviation of the
cluster radial extent with increasing $\alpha$. Left panel
corresponds to $a_{1x}$ ($m=0)$, middle to $a_{1y}$ ($m=1)$ and
right to $a_{1z}$. \label{fig:extent}}
 \end{figure}
\clearpage

Of course, as the the mean cluster extent gets increased further
and we reach passed the cluster gas extent radii we should also
observe a decrease in the measured dipole. To check this we have
run our pipeline with the net cluster angular extent increased to
1, 2 and 3 degree radii. At the $3^\circ$ radii the cluster
catalog occupies a significant fraction of the available sky,
$\simeq 35\%$, so at larger radii the clusters overlapping would
become significant. We observe that the dipole, in all $z$-bins
where we have a statistically significant measurement, indeed
decreases with the increasing mean extent for the apertures with
$\alpha \gsim 6$. This is shown in Fig. \ref{fig:extent}. It is
interesting to note that as we increase the extent further we may
be detecting signs of the two other components of the dipole
($x,z)$, as testified by the small scatter among the mean dipole
from all the eight DA's. This is because the noise, reflected in
the scatter among the eight DA's, decreases faster than the
dilution factor in the measured dipole. However, it would be
difficult to interpret these results with the current version of
our X-ray catalog.

\section{Calibration: translating $\mu$K into km/sec}
\label{calibration}

In order to translate the CMB dipole in $\mu$K into the amplitude
of $V_{\rm bulk}$ in km/sec, we proceed as follows. First, we
verified that our catalog reproduces accurately the measured TSZ
properties of the measured CMB parameters (see also Sec.
\ref{tsz}). Table 4 compares the directly determined TSZ
contributions in the redshift bins where we have a statistically
significant detection of the dipole with those determined from the
parameters in the catalog. The latter is determined as follows:
for each cluster we construct a TSZ map in each WMAP channel using
the catalog values for the electron density, core radius, X-ray
temperature and total extent, and assuming $\beta=2/3$. These maps
are then filtered using the filters shown in Fig.
\ref{fig:filters} and coadded using the weights used in the main
pipeline. As a consistency check we determine the gas profile
using two independent methods: (a) fitting a $\beta$-profile
directly to the RASS X-ray data, and (b) using an empirical
relationship between the core radius and X-ray luminosity. The
quantities derived from the catalog should have the same
uncertainties (generated by the CMB maps noise etc) as those
measured directly and for brevity are not shown. The table shows
that there is good agreement between the directly measured TSZ
component and that derived using the X-ray cluster catalog for
$\theta_{\rm
  SZ}=\theta_{\rm X-ray}$. The two sets of numbers mostly overlap at 1-$\sigma$
level and always overlap at 2-$\sigma$. To further check that the
agreement is not accidental, we have generated a test catalog
randomly assigning the various cluster parameters from different
clusters. The agreement completely disappears and the two sets of
numbers become different by factors of $\sim 2-3$.

\clearpage
\begin{deluxetable}{cc|cccc|cccc}
 \tabletypesize{\scriptsize}
 \tablewidth{0pt}
 \tablecaption{TSZ component in filtered maps: observed and modelled.} \tablehead{
 \colhead{(1)} & \colhead{(2)} & \multicolumn{8}{c}{(3)} \\
 \cline{3-10}
\colhead{} &
 \colhead{CMB maps} &
 \multicolumn{8}{c}{TSZ estimate using catalogs: (a) $|$ (b)} \\

 \colhead{$z\leq$} &
 \colhead{$\langle \Delta T\rangle$} &
 \colhead{$\langle \Delta T\rangle$} &
 \colhead{$\frac{a_{1,{\rm x}}}{\langle \Delta T\rangle}$} &
 \colhead{$\frac{a_{1,{\rm y}}}{\langle \Delta T\rangle}$} &
 \colhead{$\frac{a_{1,{\rm z}}}{\langle \Delta T\rangle}$} &
 \colhead{$\langle \Delta T\rangle$} &
 \colhead{$\frac{a_{1,{\rm x}}}{\langle \Delta T\rangle}$} &
 \colhead{$\frac{a_{1,{\rm y}}}{\langle \Delta T\rangle}$} &
 \colhead{$\frac{a_{1,{\rm z}}}{\langle \Delta T\rangle}$}  \\

}
  \startdata
  & $\mu$K & $\mu$K & & & & $\mu$K & & & \\
  \hline
0.05 & $-4.5 \pm 1.3$ & $-5.3$ & 0.3 & $-0.2$ & $-0.2$ & $-5.6$ & 0.2 & $-0.2$ & $-0.2$ \\
0.06 & $-6.8 \pm 1.1$ & $-5.7$ & 0.3 & $-0.3$ & $-0.2$  & $-6.1$ & 0.3 & $-0.2$ & $-0.2$ \\
0.08 & $-7.5 \pm 1.0$ & $-6.2$ & 0.2 & $-0.0$ & $-0.1$  & $-6.7$ & 0.2 & $-0.1$ & $-0.1$ \\
0.12 & $-7.6 \pm 0.9$ & $-7.5$ & 0.1 & $0.0$ & $-0.2$ & $-7.8$ & 0.1 & 0.1 & $-0.2$ \\
0.16 & $-7.3 \pm 0.8$ & $-7.9$ & 0.2 & $-0.1$ & $-0.1$ & $-8.6$ & 0.2 & $-0.0$ & $-0.1$ \\
0.20 & $-7.4 \pm 0.8$ & $-8.8$ & 0.1 & $-0.0$ & $-0.1$ & $-9.75$ & 0.1 & $-0.0$ & $-0.1$ \\
0.30 & $-7.9 \pm 0.8$ & $-11.$ & 0.2 & $-0.0$ & $-0.0$ & $-11.9$ & 0.2 & $-0.1$ & $-0.0$ \\
\enddata
\tablecomments{Column (1) is the redshift bin of the clusters and
(2) shows the observed temperature decrement in the WMAP data for
$\theta_{\rm SZ}=\theta_{\rm X-ray}$ in each of the bins. Columns
(3) correspond to the TSZ temperature decrement and its relative
dipole calculated from the X-ray catalog data. In columns (a) and
(b) the TSZ temperature decrement is calculated using cluster
parameters derived from our best-fit $\beta$-model to the RASS
data and the empirical relationship of Reiprich \& B\"{o}hringer
(1999), respectively. CMB temperature decrements are in $\mu$K.}
\end{deluxetable}
\clearpage

Thus the cluster properties in the catalog are determined
reasonably well to estimate the translation factor between the CMB
dipole amplitude and the bulk flow velocity. To account for the
attenuation of the clusters' $\tau$ values by both the beam and
the filter, we convolve the gas profile of each cluster with the
beam and the filter shown in Fig. \ref{fig:filters} over the WMAP
pixels associated with it. Each cluster is given a bulk flow
motion of 100 km/sec in the direction listed in Table 2, so that
each pixel of the $i$-th cluster has $\delta T=T_{\rm
CMB}\tau_i(\theta) V_{\rm bulk}/c$, with $\theta$ being the
angular distance to the cluster center. We then compute the CMB
dipole of the resulting cluster map and average the results for
each channel map with the same weights as used in the dipole
computation. This allows us to estimate the dipole amplitude,
$C_{1,100}$, contributed by each 100 km/sec of bulk-flow. We
restrict our calculation to the central $1\theta_{\rm X-ray}$
where the $\beta$-model and NFW profiles differ by 10-30\% and
where the central values of the measured dipole are similar to the
values measured at the final aperture extent. In other words, we
assume that for each cluster all pixels measure the same velocity
(in modulus) across the sky, so the calibration constant, measured
from any subset of pixels is the same, irrespective of the signal
(in $\mu$K) measured at their location.

The results are shown in the last column of Table 2 for the
central values of the direction of the measured flow; varying the
direction within the uncertainties of $(l,b)$ shown in Table~2
changes the numbers by at most a few percent. A bulk flow of 100
km/sec thus leads to $\sqrt{C_1}\simeq 0.8\mu$K for unfiltered
clusters; this corresponds to an average optical depth of our
cluster sample of $\langle \tau \rangle \simeq 10^{-3}$ consistent
with what is expected for a typical galaxy cluster. Filtering
reduces the effective $\tau$ by a factor of $\simeq 3$. As
mentioned above, since a $\beta$-model provides a poor fit to the
measured TSZ component outside the estimated values of
$\theta_{\rm X-ray}$ \cite{paper2}, we compute $C_{1,100}$ with
the total extent assumed to be $\theta_{\rm X-ray}$ where the
central value of the bulk-flow dipole has approximately the same
value as at the final aperture of $\min[6\theta_{\rm
  X-ray},30^\prime]$. Owing to the large size of our cluster sample ($N_{\rm cl}
\sim$130-675), the random uncertainties in the estimated values of
$C_{1,100}$ should be small, but we cannot exclude a systematic
offset related to selection biases affecting our cluster catalog
at high redshift.  Any such offset, if present, will become
quantifiable with the next version of our X-ray cluster catalog
(in preparation) which will use the empirically established SZ
profile \cite{paper2} rather than the currently used $\beta$-model
to parameterize the cluster gas profile. The good agreement
between the various TSZ-related quantities shown in Table 4 for
$\theta_{\rm SZ}=\theta_{\rm X-ray}$ and the observed values for
both unfiltered \cite{paper2} and filtered maps suggests, however,
that these systematic uncertainties are not likely to be high. We
also note that they only affect the accuracy of the determination
of the amplitude of the bulk flow, but cannot put its existence
into doubt which is established from the CMB dipole detected at
the cluster locations. Since the filtering effectively removes the
profile outside, approximately, a few arcmin (see Fig.
\ref{fig:filters}), it removes a more substantial amount of power
in the $\beta$-model when the cluster SZ extent is increased
beyond $\theta_{\rm X-ray}$, than in the steeper profile measured
by us \cite{paper2}. Therefore, the effective $\tau$ is possibly
underestimated by using a $\beta$-model. Nevertheless, the
calibration factor {\it cannot exceed} $\sqrt{C_{1,100}}\simeq 0.8
\mu$K given by that of the unfiltered clusters, so the measured
flow has bulk velocity of at least a few hundred km/sec
independently of scale out to at least $\gsim 300 h^{-1}$Mpc. The
above number for the calibration is {\it lowered} by filtering.
Filtering removes somewhat more power in the NFW clusters than in
the $\beta$-model, so the value of $\sqrt{C_{1,100}}=0.3 \mu$K for
filtered clusters in Table 2, is a firm lower limit. At the same
time, the central dipole value there is more-or-less the same as
for larger apertures. Fig. \ref{fig:systematics} shows that
geometrical considerations do not introduce more that a few
percent in the calibration constant.

While the above already limits calibration to a relatively narrow
range, a more accurate determination of $C_{1,100}$ would require
an adequate knowledge, not yet available, of the NFW profile of
each individual cluster. It is not sufficient to know the average
profile of the cluster population (AKKE). Filtering acts
differently on the NFW-type clusters depending on their angular
extent and concentration parameter, i.e., the filtered mean
profile is not the same as the mean of all filtered profiles.
However, since $C_{1,100}$ was computed using the central pixels,
the region where the filter preserves the signal most  and where
both profiles differ less, we believe that our estimate of
$C_{1,100}\simeq 0.3\mu$K is fairly accurate, at least in the
sense that our overall cosmological interpretation holds within
the remaining uncertainties and it is fairly independent of the
cluster sub-samples in Table 2.

\section{Future prospects}
\label{future}

The noise of our measurement of the dipole at $1.8 (N_{\rm
cl}/100)^{-1/2}\mu$K with three-year WMAP data is in good
agreement with the expectations of \cite{kab}. The uncertainties
in our measurement are dominated by the instrument noise and
should thus decrease toward the end of the 8-year WMAP mission by
a factor of $\sqrt{8/3}\simeq 1.6$. This should enable us to
measure the flows with an accuracy for individual $a_{1m}$ values
of $\simeq 1$ to $\simeq 0.25 \mu$K for $z\leq 0.03$ and $z\leq
0.3$, improving the accuracy of the measurement and perhaps
uncovering the flows at lower $z$ and the currently undetermined
components of the dipole. Particularly useful in the future would
be to make such measurement at around 217 GHz, where the TSZ
component vanishes, and at larger frequencies where it changes
sign. This could be achievable with the planned ESA-led Planck CMB
mission (http://www.rssd.esa.int/planck).

After this project was completed, the WMAP mission has released
its 5-year integration data. The data ave lower noise than the
3-year integrations used here. We will report the full results
from the 5-year data analysis (and extended X-ray cluster catalog
- see next paragraph) in separate publications after the full work
is completed. Suffice is to say here that our preliminary analysis
of the 5-year CMB maps gives results in full agreement with this
paper. However, because the new CMB mask of the 5-year data
release, KQ75, is somewhat different and larger than the KP0 mask
of the 3-year data, fewer clusters can enter the final analysis
and the reduction in errors seems less than $\sqrt{5/3}=1.3$. This
will be improved with a new expanded cluster catalog we are
developing now as described in the following paragraph.

Another obvious avenue toward improving this measurement goes
through an increased cluster sample. Since X-ray selection is
critical to ensure that all systems selected are indeed
gravitationally bound, and since all-sky (or near-all-sky)
coverage is crucial to ensure unbiased sampling of the dipole
field, the database of choice for this purpose remains the ROSAT
All-Sky Survey (RASS). The cluster sample used in our present work
can be straightforwardly extended by adopting a lower X-ray flux
limit. While this will not result in a noticeable increase of our
sample at low redshift (at much lower X-ray fluxes than used by us
here we would begin to select very poor galaxy groups and even
individual galaxies), tremendous statistical gains can still be
made at redshifts greater than, say, 0.15 where our present flux
limit excludes all but the most X-ray luminous systems. We
therefore are working to extend to the whole sky the approach
successfully taken by the MACS project (Ebeling et al. 2001,
2007), i.e. to identify clusters in the RASS data down to detect
fluxes of $1\times 10^{-12}$ erg cm$^{-2}$ s$^{-1}$ (0.1--2.4
keV), thereby extending our study to redshifts approaching 0.7. We
note that the poor photon statistics of the RASS (a detection at
such low fluxes consists often of no more than 20 X-ray photons)
are irrelevant for our purposes as long as the cluster nature of
the X-ray source can be unambiguously confirmed. MACS has
demonstrated that this is possible, specifically at high redshift,
by means of imaging follow-up observations at optical wavelengths.
(Since we recover the CMB dipole which exists at high significance
level only at the CMB pixels associated with X-ray clusters even
adding a small fraction of CMB pixels not associated with true
clusters can only decrease the statistical significance of the
results, rather than introduce bias). Clusters at $z>0.1$ are
essentially unresolved in the RASS, and are most definitely
unresolved in the WMAP data, meaning that both surveys are
sensitive only to the integrated cluster signal which is
independent of the exact shape of the X-ray emission (radial
surface-brightness profile, general morphology). The compilation
of a well defined, RASS-selected, all-sky cluster sample following
the MACS selection criteria is currently done by us for this
project in conjunction with longer integration WMAP data.

We are currently developing ways to improve our calibration of
$C_{1,100}$ using a directly fit NFW profile for our catalog
clusters. In AKKE we have measured the average NFW of our cluster
sample. The poor resolution of WMAP data, the amplitude of the
intrinsic CMB signal compared with the TSZ contribution and the
limited frequency range of WMAP radiometers may limit the ability
to estimate the NFW for each individual cluster in our sample
using the available CMB data. The PLANCK mission, with its large
frequency coverage will permit to estimate those parameters with
enough precision for the purposes of this project. Although our
calibration uncertainty is unlikely to exceed $\sim 20-30\%$, the
newly constructed catalog should narrow down these systematic
effects even more.

Further improvements can be done by specifically designing more
optimal filtering schemes to isolate specifically the
contributions from the clusters of galaxies to CMB anisotropies.
Here care is required. Our filter is based on the data and the
actual realization of the noise. It, eq. \ref{eq:filter}, is
specifically designed to eliminate the cosmological fluctuations
in a given (random and channel-specific) noise realization, which
is done efficiently enough as our results show, because the power
spectrum of the largest contributor to the dipole, the
cosmological CMB fluctuations, is known with high accuracy. If one
uses more theoretical filters, e.g. to isolate the SZ component of
the power spectrum, the latter must be known with high accuracy
(say, at least as high as the $\Lambda$CDM CMB power spectrum) and
it must be known with high accuracy for our catalog clusters.
Furthermore, Wiener-type filters do not preserve power and
different filters remove different amounts of it. Thus the
additional filter-specific issues would be the different
calibration procedures and the different monopole (from TSZ) in
the residual maps.

\section{Summary}
\label{summary}

We now summarize the main conclusions from this study:

$\bullet$ Our measurements indicate the existence of the residual
CMB dipole evaluated over the CMB pixels associated with the hot
SZ producing gas in clusters of galaxies. The dipole is measured
at high-signifance level ($\sim 8\sigma$ in the outer bins) and
persists out the limit of our cluster catalog $z_{\rm
median}\simeq 0.1$. Its direction is not far off the direction of
the "global CMB dipole" measured from the entire unprocessed maps.

$\bullet$ We show with detailed simulation that the CMB mask
and/or cluster sample discreteness induced cross-talk effects are
negligible and cannot mimic the measured dipole.

$\bullet$ The dipole originates exclusively at the cluster pixels
and, hence, cannot be produced by foregrounds or instrument noise.
It must originate from the CMB photons that have passed through
the hot gas in the catalog clusters.

$\bullet$ We prove that the signal arises from the hot SZ
producing cluster gas because we demonstrate that in the
unfiltered CMB maps there remains statistically significant
temperature {\it decrement} as expected from the TSZ effect. Its
profile is consistent with the NFW profile out the largest
aperture where we still detect hot gas ($\sim 30^\prime$). At
larger radii the dipole begins to decrease as expected.

$\bullet$ In the filtered maps, designed to reduce the
cosmological CMB fluctuations, the dipole is isolated
simultaneously as the monopole component vanishes. This proves
that its origin lies in the KSZ component. The monopole vanishes
(within the noise) because for the NFW profile the gas in
hydrostatic equilibrium must have a strong decrease in the X-ray
temperature in the outer parts. This decrease is consistent with
the available direct X-ray measurements, but more importantly is
demonstrated empirically in AKKE.

$\bullet$ With the current cluster catalog we determine that the
amplitude of the dipole corresponds to bulk flow of 600-1000
km/sec. This conversion factor, $C_{\rm 1,100}$, may however have
some systematic offset related to our current cluster modelling.
However, this possible uncertainty only affect the amplitude of
the motion, not its coherence scale or existence.

$\bullet$ The cosmological implications are discussed in
Kashlinsky et al (2008). We show there that the concordance
$\Lambda$CDM model cannot account for this motion at many standard
deviations. Instead, it is possible that this motion extends all
the way to the current cosmological horizon and may originate from
the tilt across the observable Universe from far away
pre-inflationary inhomogeneities \cite{ktf,turner}.

This work is supported by the NASA ADP grant NNG04G089G in the USA
(PI - A. Kashlinsky) and by the Ministerio de Educaci\'on y
Ciencia and the ''Junta de Castilla y Le\'on'' in Spain
(FIS2006-05319, PR2005-0359 and SA010C05, PI - F.
Atrio-Barandela). We thank Gary Hinshaw for useful information
regarding the WMAP data specifics. FAB thanks the University of
Pennsylvania for its hospitality when part of this work was
carried out. We thank Carlos Hernandez-Monteagudo for spotting a
technical correction in the SZ energy distribution, eq. 4 and Fig.
7.

{\it NOTE ADDED IN PROOF}: Our results have received recently
additional support from an independent study by Watkins, Feldman
and Hudson (arXiv:0809.4041; 2009, MNRAS, 392, 743) . The Watkins
et al. study compiled all major peculiar velocity surveys to date
to determine bulk flows within a 100 $h^{-1}$Mpc sphere. Although
the scales involved are much smaller than, and the method
completely different from ours, Watson et al find that the
galaxies within a $\sim 50-100 h^{-1}$Mpc sphere are moving at a
significant velocity in the same direction as found in our work.
The amplitude of their motion, at $\sim400-500$ km/sec, appears
somewhat smaller, but still overlaps within $<$ 2 standard
deviations with our velocity assuming the calibration above. We
anticipate that recalibrating the cluster sample as described in
Sec. 8 will further decrease the difference between the measured
velocity amplitudes.





\begin{thebibliography}{3}
\bibitem [Aghanim et al 2008]{aghanim}{Aghanim, N., Majumdar, S. \& Silk, J. 2008, Rep. Prog. Phys., 71,
066902}
\bibitem [Atrio-Barandela et al 2008]{paper2}{Atrio-Barandela, F., Kashlinsky, A.,
Kocevski, D. \& Ebeling, H. 2008, Ap.J. (Letters), 675, L57.
(AKKE)}
\bibitem [Birkinshaw 1999]{birkinshaw}{Birkinshaw, M. 1999,
Ph.Rep., 310, 97}
\bibitem [Bohringer et al 2004]{bohringer}{B\"{o}hringer, et al. 2004, A\&A, 425,
367}
 \bibitem [Borgani et al 2004]{borgani}{Borgani, S. et al 2004, MNRAS, 348,
1078}
 \bibitem [Cavaliere \& Fusco-Femiano 1976]{beta-model}{Cavaliere, A. \& Fusco-Femiano, R. 1976, A\&A, 49,
 137}
\bibitem [Carlstrom et al 2002]{carlstrom}{Carlstrom, J.E.,
Holder, G.P. \& Reese, E.D. 2002, ARA\&A, 40, 643}
\bibitem [Courteau et al 2000]{courteau}{Courteau, S. et al 2000, Ap.J., 544, 636}
\bibitem [Djorgovski \& Davis 1987]{djorgovski}{Djorgovski, S. \&
Davis, M. 1987, Ap.J., 313, 59}
\bibitem [Dressler et al 1987]{7s-di}{ Dressler et al 1987, Ap.J., 313,
42}
\bibitem [Ebeling et al 1998]{ebeling1}{Ebeling, H., Edge, A.C., B\"{o}hringer, H., Allen, S.W., Crawford,
C.S., Fabian, A.C., Voges, W., \& Huchra, J.P. 1998, MNRAS,  301,
881}
\bibitem [Ebeling et al 2000]{ebeling2}{Ebeling, H., Edge A.C., Allen S.W., Crawford C.S., Fabian A.C., \&
Huchra J.P. 2000, MNRAS, 318 333}
\bibitem [Ebeling et al 2001]{ebeling2001}{Ebeling, H. et al 2001, TBD}
\bibitem [Ebeling et al 2002]{ebeling3}{Ebeling, H.,  Mullis, C.R., \& Tully R.B. 2002, ApJ, 580,
774}
\bibitem [Ebeling et al 2007]{ebeling2001}{Ebeling, H. et al 2007, TBD}
 \bibitem [Gorski et al 2005]{healpix}{Gorski, K. et al 2005, Ap.J., 622,
 759}
 \bibitem [Jones \& Forman 1984]{jones-forman}{Jones, C. \& Forman, W. 1984, ApJ, 276,
 38}
 \bibitem [Henry \& Hendriksen 1986]{henry}{Henry, J.P. \& Henriksen, M.J. 1986, Ap.J., 301,
 689}
 \bibitem [Hinshaw et al 2007]{hinshaw}{Hinshaw, G. et al 2007, Ap.J., 170, 288}
 \bibitem [Holzapfel et al 1997]{holzapfel}{Holzapfel, W.L. et al
1997, Ap.J., 479, 17}
\bibitem [Hudson \& Ebeling 1997]{hudson-ebeling}{Hudson, M.J. \& Ebeling, H.
1997, Ap.J., 479, 621}
\bibitem [Hudson et al 1999]{hudson}{Hudson, M.J. et al 1999
Ap.J., 512, L79}
\bibitem [Kashlinsky 1988]{k88}{Kashlinsky, A. 1988, Ap.J., 331,
L1}
\bibitem [Kashlinsky \& Atrio-Barandela 2000]{kab}{Kashlinsky, A. \& Atrio-Barandela, F. 2000, Ap.J., 536,
L67}
 \bibitem [Kashlinsky \& Jones 1991]{kashlinsky-jones}{Kashlinsky,
A. \& Jones, B.J.T. 1991, Nature, 349, 753}
\bibitem [Kashlinsky et al 1994]{ktf}{Kashlinsky, A., Tkachev,
I., Frieman, J. 1994, Phys. Rev. Lett., 73, 1582}
\bibitem [Kashlinsky et al 2008]{longpaper}{Kashlinsky, A.,
Atrio-Barandela, F., Kocevski, D. \& Ebeling, H. 2008,
Ap.J.(Letters), 686, L49.}
 \bibitem [Kocevski \& Ebeling 2006]{kocevski2}{Kocevski, D.D. \& Ebeling, H. 2006, ApJ, 645,
1043}
\bibitem [Kocevski et al 2007]{kocevski1}{Kocevski, D.D., Ebeling, H., Mullis, C.R., \& R.B. Tully, 2007, ApJ,
\emph{in press}}
\bibitem [Kocevski et al 2004]{kocevski3}{Kocevski, D.D., Mullis, C.R., \& Ebeling, H. 2004, ApJ, 608,
721}
 \bibitem [Komatsu \& Seljak 2001]{komatsu}{Komatsu, E. \& Seljak, U. 2001, MNRAS,
1353}
\bibitem [Lauer \& Postman 1994]{lauer-postman}{Lauer, T. R. \&
Postman, M. 1994, Ap.J., 425, 418}
 \bibitem [Lynden-Bell et al 1988]{7s-motion}{Lynden-Bell, D. et al
1988, Ap.J., 326, 19}
\bibitem [Mathewson et al 1992]{mathewson}{Mathewson, D.S., Ford,
V.L. \& Buchhorn, M. 1992, Ap.J., 389, L5}
\bibitem [Navarro et al 1996]{nfw}{Navarro, J.F., Frenk, C.S. \& White, S.D.M. 1996, ApJ, 462,
 563}
\bibitem [Nolta et al 2008]{nolta}{Nolta, M.R. et al 2008, ApJS,
in press}
\bibitem [Phillips 1995]{phillips}{Phillips, P.R. 1995, Ap.J., 455, 419}
 \bibitem [Pratt et al 2007]{pratt}{Pratt, G.W. et al 2007, Astron. Astrophys. 461,
 71}
 \bibitem [Press et al 1986]{recipes}{Press, W.H., Flannery, B.P., Teukolsky, S.A. \& Vetterling, W.T.
1986, {\it Numerical Recipes}, Cambridge University Press}
 \bibitem [Raymond \& Smith 1977]{raymond-smith}{Raymond, J.C. \& Smith, B.W. 1977, Ap. J. Suppl., 35,
 419}
 \bibitem [Rees \& Sciama 1968]{rees-sciama}{Rees, M.J., \& Sciama, D. 1968, Nature, 511,
 611}
\bibitem [Reiprich \& Boehringer 1999]{reiprich}{Reiprich, T. H., \& Boehringer, H. 1999, Astron. Nachr., 320,
 296}
\bibitem [Riess et al 1997]{riess}{Riess, A., Davis, M., Baker, J. \& Kirshner, R.P. 1997, Ap.J., 488, L1}
 \bibitem [Rubin et al 1976]{rubin-ford}{Rubin, V., Roberts, M., Thonnard \& Ford, W.K. 1976, A.J., 81,719}
\bibitem [Scaramella et al 1991]{scaramella}{Scaramella, R., Vettolani, G., \& Zamorani, G. 1991, ApJ, 376,
L1}
 \bibitem [Stebbins 1997]{stebbins}{Stebbins, A. 1997,
 astro-ph/9705178}
\bibitem [Strauss \& Willick 1995]{strauss-willick}{Strauss, M. \&
Willick, J.A. 1995, Phys. Rep., 261, 271}
\bibitem [Turner 1991]{turner}{Turner, M. S. 1991, Phys.Rev., D44,
3737}
 \bibitem [Voges et al 1999]{voges}{Voges, W., et al. 1999, A\&AS, 349,
 389}
 \bibitem [White et al 1997]{white}{ White D.A., Jones C., Forman W. 1997, MNRAS, 292,
 419}
\bibitem [Willick 1999]{willick99}{Willick, J.A. 1999, Ap.J., 522,
647}
\bibitem [Willick 2000]{willick}{Willick, J.A. 2000,
astro-ph/0003232, in Proceedings of the XXXVth Rencontres de
Moriond: Energy Densities in the Universe}

\end{thebibliography}
\end{document}